\documentclass[prd,aps,twocolumn,superscriptaddress,nofootinbib,floatfix]{revtex4}

\usepackage{amsmath,graphicx,epsfig}
\usepackage{epstopdf}
\usepackage{euscript}
\usepackage{amsfonts}
\usepackage{amssymb}
\usepackage{float}
\usepackage[colorlinks = true,
            linkcolor = blue,
            urlcolor  = blue,
            citecolor = blue,
            anchorcolor = blue]{hyperref}
\def\prn#1{{\left(#1\right)}}

\def\abrk#1{{\langle#1\rangle}}

\def\bra#1{{\langle#1|}}

\def\cg(#1,#2)(#3,#4)(#5,#6){\bra{#1,#2,#3,#4}#5,#6\rangle}

\def\ts#1{{_{\mbox{\scriptsize #1}}}}

\def\threej(#1,#2)(#3,#4)(#5,#6){\begin{pmatrix}#1&#3&#5\\#2&#4&#6\end{pmatrix}}
\def\sixj(#1,#2,#3)(#4,#5,#6){\begin{Bmatrix}#1&#2&#3\\#4&#5&#6\end{Bmatrix}}
\def\ninej(#1,#2,#3)(#4,#5,#6)(#7,#8,#9){\begin{Bmatrix}#1&#2&#3\\#4&#5&#6\\#7&#8&#9\end{Bmatrix}}

\def\bs{\boldsymbol}
\def\mc{\mathcal}

\def\EE[#1]{\mathbf{E} [#1] }
\def\VV[#1]{\mathbf{Var} [#1] }

\newlength{\defbaselineskip}
\newcommand{\setlinespacing}[1]%
           {\setlength{\baselineskip}{#1 \defbaselineskip}}

\usepackage{color}

\usepackage[]{ezedits}

\defineEdit{NLF}{\color[rgb]{.9,.35,.0}}{\color[rgb]{1,.5,.0}}
\defineEdit{HMR}{\color[rgb]{.5,.4,.9}}{\color{blue}}
\defineEdit{DK}{\color[rgb]{.7,.9,.8}}{\color[rgb]{0,.5,.3}}
\defineEdit{AW}{\color{pink}}{\color[rgb]{1., 0.5, 0.6}}
\defineEdit{AB}{\color{purple}}{\color[rgb]{.26, .50, .33}}
\defineEdit{DB}{\color{blue}}{\color[rgb]{0, 0, 1}}
\defineEdit{JAS}{\color[rgb]{.9,.7,.4}}{\color[rgb]{1, .6, 0}}

\newcommand{\cut}[1]{}

\begin{document}

\title{Intensity interferometry for ultralight bosonic dark matter detection}

\author{Hector~Masia-Roig}
\email{hemasiar@uni-mainz.de}
\affiliation{Johannes Gutenberg-Universit\"at Mainz, 55128 Mainz, Germany}
\affiliation{Helmholtz-Institut Mainz, GSI Helmholtzzentrum f{\"u}r Schwerionenforschung, 55128 Mainz, Germany}

\author{Nataniel~L.~Figueroa}
\email{figueroa@uni-mainz.de}
\affiliation{Johannes Gutenberg-Universit\"at Mainz, 55128 Mainz, Germany}
\affiliation{Helmholtz-Institut Mainz, GSI Helmholtzzentrum f{\"u}r Schwerionenforschung, 55128 Mainz, Germany}

\author{Ariday~Bordon}
\email{aridaybordon@gmail.com}
\affiliation{Johannes Gutenberg-Universit\"at Mainz, 55128 Mainz, Germany}
\affiliation{Helmholtz-Institut Mainz, GSI Helmholtzzentrum f{\"u}r Schwerionenforschung, 55128 Mainz, Germany}

\author{Joseph~A.~Smiga}
\affiliation{Johannes Gutenberg-Universit\"at Mainz, 55128 Mainz, Germany}
\affiliation{Helmholtz-Institut Mainz, GSI Helmholtzzentrum f{\"u}r Schwerionenforschung, 55128 Mainz, Germany}

\author{Yevgeny~V.~Stadnik}
\affiliation{School of Physics, The University of Sydney, New South Wales 2006, Australia}

\author{Dmitry~Budker}
\affiliation{Johannes Gutenberg-Universit\"at Mainz, 55128 Mainz, Germany}
\affiliation{Helmholtz-Institut Mainz, GSI Helmholtzzentrum f{\"u}r Schwerionenforschung, 55128 Mainz, Germany}
\affiliation{Department of Physics, University of California at Berkeley, Berkeley, California 94720-7300, USA}

\author{Gary~P.~Centers}
\affiliation{Johannes Gutenberg-Universit\"at Mainz, 55128 Mainz, Germany}
\affiliation{Helmholtz-Institut Mainz, GSI Helmholtzzentrum f{\"u}r Schwerionenforschung, 55128 Mainz, Germany}



\author{Alexander~V.~Gramolin}
\affiliation{Department of Physics, Boston University, Boston, MA 02215, USA}

\author{Paul~S.~Hamilton}
\affiliation{Department of Physics and Astronomy, University of California, Los Angeles, California 90095, USA}

\author{Sami~Khamis}
\affiliation{Department of Physics and Astronomy, University of California, Los Angeles, California 90095, USA}

\author{Christopher~A.~Palm}
\affiliation{Department of Physics, California State University -- East Bay, Hayward, California 94542-3084, USA}


\author{Szymon~Pustelny}
\affiliation{Institute of Physics, Jagiellonian University, 30-059 Krak\'ow, Poland}



\author{Alexander~O.~Sushkov}
\affiliation{Department of Physics, Boston University, Boston, MA 02215, USA}
\affiliation{Department of Electrical and Computer Engineering, Boston University, Boston, MA 02215, USA}
\affiliation{Photonics Center, Boston University, Boston, MA 02215, USA}

\author{Arne~Wickenbrock}
\affiliation{Johannes Gutenberg-Universit\"at Mainz, 55128 Mainz, Germany}
\affiliation{Helmholtz-Institut Mainz, GSI Helmholtzzentrum f{\"u}r Schwerionenforschung, 55128 Mainz, Germany}

\author{Derek~F.~Jackson Kimball}
\email{derek.jacksonkimball@csueastbay.edu}
\affiliation{Department of Physics, California State University -- East Bay, Hayward, California 94542-3084, USA}



\begin{abstract}
Ultralight bosonic dark matter (UBDM) can be described by a classical wave-like field oscillating near the Compton frequency of the bosons. If a measurement scheme for the direct detection of UBDM interactions is sensitive to a signature quadratic in the field, then there is a near-zero-frequency (dc) component of the signal. Thus, a detector with a given finite bandwidth can be used to search for bosons with Compton frequencies many orders of magnitude larger than its bandwidth. This opens the possibility of a detection scheme analogous to Hanbury Brown and Twiss intensity interferometry. Assuming that the UBDM is virialized in the galactic gravitational potential, the random velocities produce slight deviations from the Compton frequency. These result in stochastic fluctuations of the intensity on a time scale determined by the spread in kinetic energies. In order to mitigate ubiquitous local low-frequency noise, a network of sensors can be used to search for the stochastic intensity fluctuations by measuring cross-correlation between the sensors. This method is inherently broadband, since a large range of Compton frequencies will yield near-zero-frequency components within the sensor bandwidth that can be searched for simultaneously. Measurements with existing sensor networks have sufficient sensitivity to search experimentally unexplored parameter space. 
\end{abstract}

\maketitle

A wide variety of evidence suggests the existence of dark matter, an invisible substance that constitutes a large fraction of the matter in the Universe \cite{Fen10,spergel2015dark,graham2015experimental}. Despite decades of research, its microscopic nature remains unknown. A promising hypothesis is that dark matter predominantly consists of ultralight bosons with masses $m_\varphi \ll 1\,{\textrm{eV}}/c^2$, such as axions~\cite{Pre83,Abb83,Din83}, axion-like particles (ALPs)~\cite{Gra15,co2020predictions}, or hidden photons~\cite{holdom1986two,cvetivc1996implications,Okun1982}. Ultralight bosons can couple to Standard Model (SM) particles through a variety of ``portals'' \cite{graham2016dark,safronova2018search}, which have been used to search for ultralight bosonic dark matter (UBDM) --- see, for example, Refs.\,\cite{Asz01,Asz10,budker2014proposal,Sik14,chaudhuri2015radio,geraci2016sensitivity,abel2017search,wu2019search,garcon2019constraints,phipps2020exclusion,manley2020searching,crescini2020axion,braine2020extended,backes2021quantum,antypas2021probing,roussy2021experimental,Ayb21CASPErE,gramolin2021search}.

Numerous experiments looking for UBDM are based on resonant systems, which require accurate tuning to the unknown Compton frequency, $\omega_c = m_\varphi c^2 / \hbar$, in order to be sensitive to UBDM. Therefore, a time-consuming scan of tunable parameters must be performed to search a wide range of masses. However, if a search is based on the measurement of the \emph{intensity} of the UBDM, part of the signal is down-converted to near-zero frequency~\cite{stadnik2015can,hees2018violation}, regardless of the particular Compton frequency of the UBDM. 

In the commonly considered simplest version of the standard halo model (SHM) for dark matter~\cite{Zyla:2020zbs,evans2019refinement,drukier1986detecting}, the net UBDM field results from the superposition of numerous virialized bosons \cite{derevianko2018detecting}. Such a model assumes minimal self-interactions \cite{Bohua2014} and ignores possible non-virialized dark matter streams \cite{diemand2008clumps} and composite dark matter structures such as boson stars \cite{eby2016boson} or topological defects \cite{Pos13}.  The velocity dispersion of the dark matter particles in the neighborhood of our solar system is $v_0 \sim 10^{-3}c$. Consequently, the oscillation frequencies associated with the virialized bosons are Doppler-shifted. This generates a fractional shift from $\omega_c$ and a spread of frequencies of $\sim v_0^2/\prn{2c^2} \sim 5 \times 10^{-7}$. Because of the random distribution of frequencies of the UBDM, the amplitude of the net UBDM field stochastically fluctuates, as discussed in detail in Refs.~\cite{Centers2019,foster2018,hui2021wave,schive2014cosmic,turner1990periodic,o2017axion,lisanti2021stochastic,gramolin2021spectral}. We emphasize that this is an essential feature of the UBDM field in the SHM.  The characteristic time scale, $\tau_\varphi \sim  2\hbar/\prn{m_\varphi v_0^2}$,  and length scale, $\lambda_\varphi \sim \hbar / \prn{m_\varphi v_0}$, of the fluctuations depend on the mass, $m_\varphi$, and velocity dispersion, $v_0$, of the bosons~\cite{derevianko2018detecting}. In comparison to a direct measurement of the UBDM field, a measurement of the UBDM field intensity produces a frequency down-conversion of the UBDM signal to near-dc. The spectral linewidth of this near-dc signal is $\sim \! \! 10^6$~times smaller than $\omega_c$. Looking for this near-dc feature allows sensors with a limited bandwidth to probe UBDM with masses $\sim \! \! 10^6$~times larger than searching for direct field oscillations at $\omega_c$.

The frequency down-conversion discussed above occurs naturally when considering quadratic portals where the interaction with SM particles is proportional to the square of the UBDM field\,\cite{stadnik2015can,hees2018violation,Pos13,Der14,Oli08,Sta15,dailey2021quantum}. A spin-0 field $\varphi(\bs{r},t)$ can interact with SM fermions and electromagnetic fields according to the phenomenological quadratic scalar Lagrangian\,\cite{stadnik2015can}
\begin{align}
    \mc{L}_s = \hbar c \prn{\pm \frac{m_f c^2}{\Lambda_f^2} \bar{\psi}_f\psi_f \pm \frac{1}{4 \Lambda_\gamma^2} F_{\mu\nu}^2} \varphi^2(\bs{r},t)~,
    \label{eq:scalar-Lagrangian}
\end{align}
where $\Lambda_f$ and $\Lambda_\gamma$ parametrize the couplings to fermions and photons, respectively, where the $\pm$ indicates the sign of the coupling, $m_f$ is the fermion mass, $\psi_f$ is the fermion field, and $F_{\mu\nu}$ is the Faraday tensor.

The effects of the interactions described by Eq.\,\eqref{eq:scalar-Lagrangian} can be understood in terms of redefinitions of the effective fermion masses and the fine-structure constant $\alpha$\,\cite{stadnik2015can,Der14}:
\begin{align}
    m_f^{({\rm{eff}})} (\bs{r},t) & = m_f \prn{ 1 \mp \frac{\hbar c}{\Lambda_f^2} \varphi^2(\bs{r},t) }~, \label{eq:scalar-mass-variation} \\
    \alpha^{({\rm{eff}})} (\bs{r},t) & = \alpha \prn{ 1 \pm \frac{\hbar c}{\Lambda_\gamma^2} \varphi^2(\bs{r},t) }~. \label{eq:scalar-alpha-variation}
\end{align}
Variations of $m_f^{({\rm{eff}})}$ and $\alpha^{({\rm{eff}})}$ can be measured with atomic clocks or interferometers~\cite{Der14,wcislo2016experimental,roberts2017search,roberts2020search,dailey2021quantum,nagano2019axion,aiello2021constraints,badurina2020aion,Stadnik2016,Grote2019,Vermeulen2021,Sta15}, or generally by direct comparison of systems with different dependence on these fundamental constants~\cite{Antypas2019Oct,Antypas2020Apr,antypas2021probing,Oswald2021Nov,Tretiak2022Jan}.

Additionally, a spin-0 pseudoscalar (ALP) field can possess linear and/or quadratic interactions with the axial-vector current of a SM
fermion~\cite{Pos13}, $\bar{\psi}_f \gamma^\mu\gamma_\ts{5} \psi_f$,
\begin{align}
 \mc{L}_\ts{lin} = \pm\frac{1}{f_l} \bar{\psi}_f \gamma^\mu\gamma_\ts{5} \psi_f \partial_\mu \varphi(\bs{r},t)~,
 \label{eq:pseudoscalar-Lagrangian_lin}
 \\
 \mc{L}_\ts{quad} = \pm\frac{1}{f_q^2} \bar{\psi}_f \gamma^\mu\gamma_\ts{5} \psi_f \partial_\mu \varphi^2(\bs{r},t)~,
\label{eq:pseudoscalar-Lagrangian}
\end{align}
where $f_l$ and $f_q$ parameterize the linear and quadratic couplings to fermion spins,  $\gamma^\mu$ and $\gamma_\ts{5}$ are Dirac matrices. The quantum chromodynamics (QCD) axion \cite{Kim79,Shi80,Zhi80,Din81,Sik83} associated with the Peccei-Quinn solution to the strong CP-problem \cite{peccei1977cp} generally possesses the linear coupling described by Eq.\,\eqref{eq:pseudoscalar-Lagrangian_lin} (see, for example, Ref.\,\cite{Gra13}). Alternatively, in effective field theories with ALPs not associated with the QCD sector, the linear coupling term may be strongly suppressed or absent \cite{Oli08}, in which case the leading-order interaction with the axial-vector current may be the quadratic coupling of Eq.\,\eqref{eq:pseudoscalar-Lagrangian} \cite{Pos13}.
Note that, like Eq.\,\eqref{eq:pseudoscalar-Lagrangian_lin}, the description of the quadratic interaction in Eq.\,\eqref{eq:pseudoscalar-Lagrangian} is manifestly Lorentz covariant, since replacing $\varphi$ by $\varphi^2$ preserves the overall number-type Lorentz structure because $\varphi$ is a spinless field.
Effective field theories featuring an ALP-spin interaction dominated by the quadratic term have much weaker constraints on the associated coupling constant $f_q$ from astrophysics \cite{Oli08,Pos13}. Of note is the fact that while the linear interaction of Eq.\,\eqref{eq:pseudoscalar-Lagrangian_lin} is CP-conserving, the quadratic interaction in Eq.\,\eqref{eq:pseudoscalar-Lagrangian} is CP-violating, and thus could potentially play a role in baryogenesis \cite{bernreuther2002cp}. While models with interactions $\propto \varphi^2$ were originally developed phenomenologically (see Ref.\,\cite{Oli08} and references therein), string theory is an example of a fundamental theory generating quadratic interactions \cite{hinterbichler2011towards} such as those described by Eqs.\,\eqref{eq:scalar-Lagrangian} and \eqref{eq:pseudoscalar-Lagrangian}, and theoretical work in this area is ongoing. Since our proposed search method relies on signals quadratic in the ALP field, in the rest of this work we focus our attention on the quadratic coupling.

In the nonrelativistic limit, Eq.\,\eqref{eq:pseudoscalar-Lagrangian} yields the interaction Hamiltonian \cite{Pos13,dailey2021quantum} 
\begin{align}
    \mc{H}_\varphi = \mp\frac{2\hbar^2c^2}{f_q^2} \bs{S} \cdot \bs{\nabla} \varphi^2(\bs{r},t)~.
    \label{eq:quadratic-Hamiltonian}
\end{align}
Equation~\eqref{eq:quadratic-Hamiltonian} features a structure similar to that of the Zeeman Hamiltonian, $\mc{H}_Z= \gamma \bs{S} \cdot \bs{B}$, where $\gamma$ is the gyromagnetic ratio and $\bs{B}$ is a magnetic field (with the above noted difference that $\mc{H}_Z$ is CP-even while $H_\varphi$ is CP-odd). Therefore $\bs{\nabla} \varphi^2(\bs{r},t)$ couples to a fermion spin $\bs{S}$ in a manner similar to a magnetic field~\cite{Pos13}, playing the role of a ``pseudo-magnetic'' field. Thus, such pseudoscalar fields can be searched for in the spin dynamics of electrons or nuclei. While a variety of sensors could be used, here we focus on atomic (nuclear) magnetometers since these are intrinsically sensitive to Zeeman shifts~\cite{Pus13,afach2018characterization,kimball2018searching,masia-roig_analysis_2020,afach2021search,Jiang2021Dec,afach2023can}.

The effects described by Eqs.\,\eqref{eq:scalar-mass-variation},~\eqref{eq:scalar-alpha-variation},~and~\eqref{eq:quadratic-Hamiltonian} are related to $\varphi^2$, which is in turn proportional to the UBDM ``intensity'' and exhibit \mbox{near-dc} stochastic amplitude fluctuations. Although measuring low-frequency signals is technically challenging due to multiple sources of low-frequency noise, an array of independent, geographically distributed sensors will tend to have uncorrelated noise. In contrast, the slowly-changing UBDM intensity will lead to a common-mode fluctuating signal present in all detectors within a coherence length of one another, which will appear in correlations between the sensors. We advocate for the use of networks of sensors to search for these stochastic fluctuations. There are existing and proposed dark matter searches sensitive to these interactions using networks consisting of a variety of sensor types such as atomic clocks~\cite{Der14,wcislo2016experimental,roberts2017search,roberts2020search,dailey2021quantum,collaboration2021frequency}, atomic magnetometers~\cite{Pos13,Pus13,afach2018characterization,kimball2018searching,masia-roig_analysis_2020,afach2021search,fedderke2021earth,fedderke2021search,arza2021earth}, gravimeters~\cite{hu2020network,mcnally2020constraining,horowitz2020gravimeter}, laser interferometers~\cite{hall2018laser,nagano2019axion,aiello2021constraints,Stadnik2016,Grote2019,Vermeulen2021,Sta15}, and atom interferometers \cite{badurina2020aion}. The methodology described below is analogous to Hanbury Brown and Twiss intensity interferometry\,\cite{brown1956question}, and can be used to dramatically expand the range of UBDM Compton frequencies that particular sensors can probe~\cite{InterferometryNote}\nocite{derevianko2018detecting,foster2021}.

\label{sec:field-properties}


\medskip

\noindent \emph{Stochastic properties of the UBDM field.~---~}The properties of the UBDM field can be derived using the framework described in Refs.\,\cite{Centers2019,derevianko2018detecting,foster2018,hui2021wave,schive2014cosmic,turner1990periodic,o2017axion,lisanti2021stochastic,gramolin2021spectral}. In brief, assuming that UBDM does not interact with itself, each individual particle can be treated as an independent wave. In this scenario, the UBDM field is well described by a superposition of these individual waves.

Assuming that the local dark-matter energy density $\rho\ts{dm}$ is solely in the form of UBDM, such field can be modeled as the superposition of $N$ oscillators\,\footnote{The individual bosons should be modeled as quantum objects not classical fields. However, the huge occupancy numbers of each mode allows to accurately model the UBDM as a superposition of classical oscillators. For example, a boson mass of 10$^{-11}$\,eV/c$^2$ (and $\rho_\ts{dm}$ = 0.3\,GeV/cm$^3$) results in a number density of particles of $\sim$10$^{19}$\,cm$^3$. The de Broglie wavelength for particles moving with $v_0$, is $\sim10^{10}$\,cm.}
\begin{align}
    \varphi(\bs{r},t) \approx \sum_{n=1}^N \frac{\varphi_0}{\sqrt{N}} \cos\prn{ \omega_n t - \bs{k}_n \cdot \bs{r} + \theta_n }~,
    \label{eq:single-boson-field}
\end{align}
where the oscillation amplitude is given by~\cite{Gra13}
\begin{align}
    \varphi_0 = \frac{\hbar}{m_\varphi c} \sqrt{ 2 \rho\ts{dm} },
    \label{eq:single-ALP-field-amp}
\end{align}
such that the average energy density in the UBDM field comprises the totality of the local dark matter. Here, $\bs{k}_n = m_\varphi \bs{v}_n /  \hbar$ is the wave vector corresponding to $\bs{v}_n$, the velocity of the $n^\text{th}$ oscillator in the laboratory frame. The phases $\theta_n$ are randomly distributed between $0$ and $2\pi$. The oscillation frequency, $\omega_{n}$, is determined mostly by the Compton frequency, $\omega_{c}$ of the underlying ultralight boson. The kinetic energy correction to the rest energy introduces small deviations from $\omega_{c}$, so that
\begin{align}
    \omega_n \approx \omega_c \prn{ 1 + \frac{\bs{v}_n^2}{2c^2} },
\end{align}
for $v_n \ll c$. Therefore, the distribution of $\omega_n$ (and $\bs{k}_n$) is determined by the velocity distribution as observed in the laboratory frame. According to the standard halo model,~$\bs{v}_n$ follows a displaced Maxwell-Boltzmann distribution defined as
\begin{equation}
f_{\text{lab}}(\bs{v}) \approx \frac{1}{\pi^{3/2} v_0^3} \exp{\left(-\frac{(\bs{v-\bs{v}_{\text{lab}}})^2}{v_0^2}\right)}, \label{eq:f_lab_velocity}
\end{equation}
where the velocity of the lab frame is~$\bs{v}_{\text{lab}} = |\bs{v}_{\text{lab}}| \bs{\hat{z}}$, and we have ignored the escape-velocity cut-off.

Sufficiently strong self-interactions could introduce effects in the coherence properties of the UBDM field not considered in this work. However, since these interactions are expected to be relatively weak~\cite{hui2021wave,Bohua2014}, they are commonly neglected in studies of direct detection searches~\cite{foster2021,derevianko2018detecting,lisanti2021stochastic}, as in this work.  Here, our interest lies in quadratic interactions with the field, proportional to $\varphi^2(\bs{r},t)$ or $\bs{\nabla} \varphi^2(\bs{r},t)$. These quantities have two terms: one near-dc component and one fast oscillating component at $\approx 2\omega_c$. We consider sensors with a limited bandwidth $\Delta \omega \ll \omega_c$, such that the fast oscillating terms can be ignored. Therefore, only the near-dc components (denoted by the subscript~$s$) of $\varphi^2$ and $\bs{\nabla} \varphi^2$ are considered. These are written as
\begin{align}
    \varphi_s^2(\bs{r},t) &= \frac{\varphi_0^2}{2N} \sum_{n,m=1}^N \cos \prn{\omega_{nm} t -\bs{k}_{nm} \cdot \bs{r}  +\theta_{nm}}\,,\label{eq:field-squared}\\
    \bs{\nabla} \varphi_s^2(\bs{r},t) &=  \frac{\varphi_0^2}{2N} \sum_{n,m=1}^N   \bs{k}_{nm} \sin \prn{\omega_{nm} t -\bs{k}_{nm} \cdot \bs{r} +\theta_{nm}}\,, \label{eq:gradient-field-sqr} 
\end{align}
where $\omega_{nm}= \omega_{n} - \omega_{m}$, $\bs{k}_{nm} = \bs{k}_{n} - \bs{k}_{m}$, and $\theta_{nm}=\theta_{n} - \theta_{m}$. 
Since the sensors are assumed to be within the same coherence patch (such that $\Delta\bs{k} \cdot \Delta\bs{r} \approx 0$, where $\Delta \bs{k}$ is the characteristic spread of values in $\bs{k}_{nm}$ and $\Delta \bs{r}$ is the difference in the position vectors for the pair of sensors at $\boldsymbol{r}_n$ and $\boldsymbol{r}_m$), the $\bs{r}$ dependence can be neglected and we can evaluate the expressions at $\bs{r}=0$ in the following calculations~\cite{TruckNote}\nocite{grotti2018geodesy,takamoto2020test}.


The signal measured with each sensor would have a small UBDM-related component $\kappa\, \xi(t)$, where $\kappa$ accounts for the coupling of the sensor to the UBDM field, and $\xi(t)$ is either $\varphi^2_s$ (scalar interaction) or $\hat{\bs{m}} \cdot \bs{\nabla} \varphi^2_s$ (pseudoscalar gradient interaction, where $\hat{\bs{m}}$ represents the sensitive direction of the sensor). The correlations between measurable signals produced in different sensors by a UBDM field can be quantified using the degree of first-order coherence $g^{(1)}(\tau)$ for different delay times $\tau$,
\begin{equation}
    g^{(1)}(\tau) = \frac{\langle \xi(t) \xi(t+\tau) \rangle_t}{\langle \xi^2 \rangle_t}~,
    \label{eq:first-order-coherence-definition}
\end{equation}
where $\abrk{\cdots}_t$ denotes the time average. The value of $g^{(1)}(\tau)$ is a measure of the degree of correlation between $\xi(t)$ and $\xi(t+\tau)$. 

\begin{figure}[ht]
\center
\includegraphics[width=\columnwidth,trim={0 5mm 0 3mm},clip]{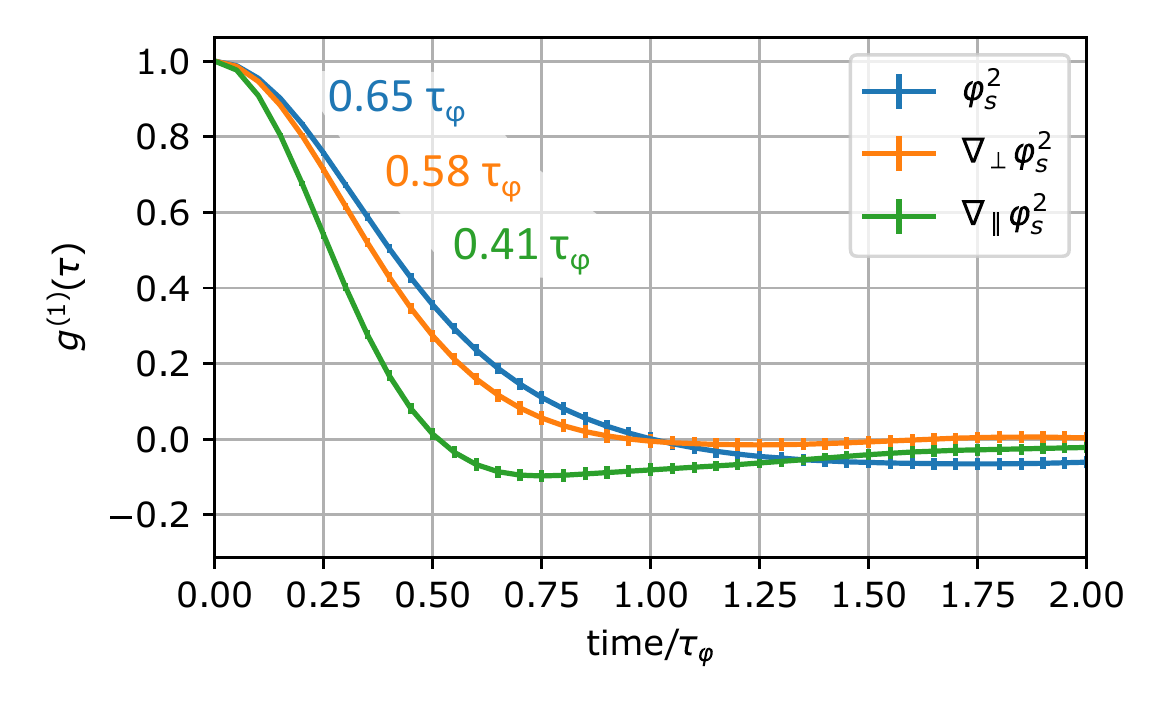}
\caption{\small{ Degree of first-order coherence as a function of delay time $\tau$ for $\varphi_s^2$ and different projections of the gradient $\bs{\nabla} \varphi_s^2$ relative to $\bs{v}_\ts{lab}$. Each curve is the result of 100~averages simulating $10^3$ particles, where the mean was subtracted. The approximate values of the coherence times are given in colors matching their respective plot traces.}}
\label{fig:g1}
\end{figure}

To illustrate the stochastic properties of $\bs{\nabla} \varphi_s^2(0,t)$ and $\varphi_s^2(0,t)$, their time evolution was simulated. Plots of $g^{(1)}(\tau)$ for $\varphi_s^2(t)$ and for projections of $\bs{\nabla} \varphi_s^2(t)$ onto parallel and perpendicular directions with respect to $\bs{v}\ts{lab}$ can be seen in Fig.\,\ref{fig:g1}. In order to numerically calculate the near-dc components of $\varphi^2$ and $\bs{\nabla} \varphi_s^2(t)$, and avoid the double summation in Eq.\,\eqref{eq:gradient-field-sqr}, it is convenient to introduce the field in complex notation
\begin{equation}
	\varphi_c(0,t) =  \frac{\varphi_0}{\sqrt{N}} \sum_{n=1}^N \exp{[i (\omega_n t + \theta_n)]} .
	\label{eq:complex-field-sum}
\end{equation}
Then the near-dc component of the field squared can be calculated using
\begin{equation}
	\varphi_s^2 = \frac{1}{2} \varphi_c \varphi_c^* .
	\label{eq:complex-scalar-field-squared}
\end{equation}
This yields a real number related to the average value of $\varphi^2$ over a cycle of the oscillation. 

Similarly, $\bs{\nabla} \varphi_s^2$ is numerically evaluated by applying the chain rule
\begin{equation}
	\begin{split}
		\bs{\nabla} \varphi_s^2(\bs{r},t)\Bigr\rvert_{\bs{r} = 0} = & \frac{i}{2} \varphi_c \frac{\varphi_0}{\sqrt{N}} \sum_{n=1}^N \bs{k}_{n} \exp{[-i (\omega_n t + \theta_n)]} \\
		& - \frac{i}{2}\varphi_c^* \frac{\varphi_0}{\sqrt{N}} \sum_{n=1}^N \bs{k}_{n} \exp{[i (\omega_n t + \theta_n)]} . \label{eq:gradient-sum-two-terms}
	\end{split}
\end{equation}

Note that by first evaluating Eq.\,\eqref{eq:complex-field-sum} to obtain $\varphi_c(0,t)$ and then evaluating Eqs.\,\eqref{eq:complex-scalar-field-squared} and \eqref{eq:gradient-sum-two-terms} to find $\varphi_s^2$ and $\bs{\nabla} \varphi_s^2$, we only evaluate sums with $\mathcal{O}(N)$ terms for $N$ oscillators. This is in contrast to the equivalent expressions presented in Eqs.\,\eqref{eq:field-squared}~and~\eqref{eq:gradient-field-sqr}, which have double sums with $\mathcal{O}(N^2)$ terms. This makes numerical calculations using Eqs.\,\eqref{eq:complex-scalar-field-squared} and \eqref{eq:gradient-sum-two-terms} considerably faster for large $N$. 

The individual wave vectors $\bs{k}_{n}$ and frequencies $\omega_{n}$ are calculated from the velocities $\bs{v}_{n}$. These velocities are drawn from the displaced Maxwell-Boltzmann distribution defined in Eq.\,\eqref{eq:f_lab_velocity}. In our simulations, we take the isotropic velocity dispersion of the local UBDM to be determined by the characteristic virial velocity $v_0\approx 220~\text{km}/\text{s}$, and $\bs{v}_{\text{lab}}$ to be dominated by the motion of the Sun in the galactic frame, $|\bs{v}_{\text{lab}}| \approx 233~\text{km}/\text{s}$. The phases $\theta_n$ are drawn from a uniform distribution spanning from $0$~to~$2\pi$.

The simulations consider $N=10^3$ oscillators evolving during $20\,\tau_\varphi$ with a time resolution of $0.05\,\tau_\varphi$, where $\tau_\varphi$ is calculated using Eq.\,\eqref{eq:coherence_time}. The number of oscillators used to model the UBDM field reflects the quantity that can be comfortably simulated with our available hardware. By repeating the simulation hundreds of times, we observe that the results converge. Additional checks confirmed that the spectral properties of the simulations matched theoretical predictions. For example, an analytical solution for $g^{(1)}(\tau)$ in the limit $|\bs{v}_{\text{lab}}| \gg v_0$ can be found in Appendix~\ref{app:analytic}, which is shown to agree with our simulations. The temporal resolution and the duration of the simulated field evolution were chosen considering plausible values for an experimental search.

After generating $\varphi_s^2$ and $\bs{\nabla} \varphi_s^2$, $g^{(1)}(\tau)$ is calculated using Eq.\,\eqref{eq:first-order-coherence-definition}. Note that the mean value of the field is subtracted so $g^{(1)}(\tau) \rightarrow 0$ for $ \tau \gg \tau_\varphi$ (see Appendix~\ref{app:search}). 

The coherence time $\tau_c$ is the characteristic time after which the correlation in the UBDM field is lost. We define the coherence time as the power-equivalent width of $g^{(1)}(\tau)$~\cite{loudon2000the,saleh2019fundamentals},
\begin{equation}
    \tau_c = \int_{-\infty}^{+\infty} |g^{(1)}(\tau)|^2 d\tau,
    \label{eq:coherence_time_def}
\end{equation}
which describes a characteristic temporal width of $g^{(1)}(\tau)$. We used this expression to quantify the coherence time in our simulations\,\footnote{For the  simulations presented here, the coherence time was obtained by integrating Eq.\eqref{eq:coherence_time_def} numerically. The integration was done over the time interval $[0,5\tau_\varphi]$, and multiplying by two in order to account for the negative segment of the range.}.

As a useful benchmark to compare our results with, we have used the coherence time $\tau_\varphi$ of the field assuming an exact Lorentzian lineshape~(with a full width at half maximum of~$\omega_c v_0^2/c^2$)~\cite{gramolin2021spectral, derevianko2018detecting},
\begin{equation}
    \tau_\varphi \approx \frac{2 \hbar}{ m_\varphi v_0^2}\,.
    \label{eq:coherence_time}
\end{equation}

Because the actual spectral lineshape describing $\varphi$ is non-Lorentzian \cite{gramolin2021spectral}, the coherence time of the field $\varphi$ derived from the simulations differs from $\tau_\varphi$. For the considered $\bs{v}_\ts{lab}$, simulations show a coherence time of~\mbox{$\approx 1.12(1) \tau_\varphi$}.

The coherence time for $\varphi^2_s$ is approximately half of that for the field $\varphi$. This is a result of the field $\varphi^2_s$ being a sum over terms depending on the difference of frequencies $\omega_{nm}$, as opposed to $\varphi$, which only contains terms depending on $\omega_n$. The probability distribution of $\omega_{nm}$ can be calculated as the convolution of the distribution of $\omega_n$ with itself. This results in a distribution for $\omega_{nm}$ that is broader. Consequently, the coherence time is shorter since $g^{(1)}(\tau)$ is given by the Fourier transform of the power spectral density of $\varphi^2_s$ (proportional to the $\omega_{nm}$ distribution), according to the Wiener–Khinchin theorem. The gradient coupling features even shorter coherence times due to the $\bs{k}_{nm}$ factor weighting the contribution of the oscillating terms. Larger $\omega_{nm}$ tends to correspond to larger $\bs{k}_{nm}$: this effectively broadens the power spectral density of $\bs{\nabla} \varphi^2_s$, leading to a shorter coherence time. This is also the reason why parallel and perpendicular components of the gradient have different coherence times, as discussed below.

\begin{figure}[ht]
\center
\includegraphics[width=\columnwidth,trim={0 5mm 0 3mm},clip]{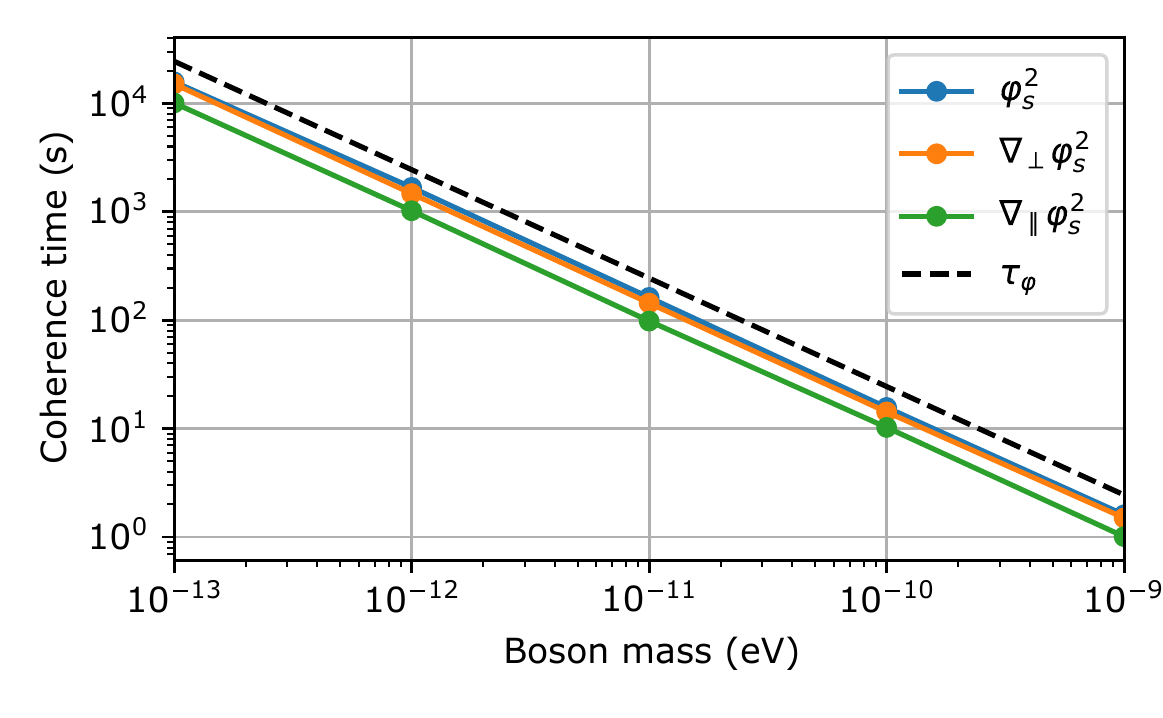}
\caption{\small{Mass dependence of the coherence time, $\tau_c$, for $\varphi_s^2$ and projections of $\bs{\nabla} \varphi_s^2$. Each point is the result of 100~averages simulating $10^3$ particles.}}
\label{Fig:massdep}
\end{figure}

The coherence time is also related to the mass of the UBDM particle as can be seen in Eq.\,\eqref{eq:coherence_time}. The relationship between the UBDM mass and the coherence time exhibited by $\varphi^2_s$ and $\bs{\nabla} \varphi^2_s$ is shown in Fig.\,\ref{Fig:massdep}. The coherence times for $\varphi^2_s$ and $\bs{\nabla} \varphi^2_s$ are proportional to the coherence time of $\varphi$, $\tau_\varphi$ as defined in Eq.\,\eqref{eq:coherence_time}. In the case of detection, this could be used to estimate the mass of the UBDM particles.

A possible method to look for UBDM is to use multi-sensor intensity interferometry to measure the cross-correlation between time-series data from different sensors.
When using pairs of geographically distributed sensors, a correlated global background field will produce a nonzero cross-correlation $g^{(1)}_{AB}(\tau)$ between sensors $A$ and $B$ proportional to $g^{(1)}(\tau)$, as discussed in Appendix~\ref{app:search}. Note that uncorrelated noise in the sensors will reduce the expected value of $g^{(1)}_{AB}(0)$ in the presence of a UBDM signal, making it smaller than the maximum value of one (see Appendix~\ref{app:search}). 
In the case of the gradient coupling, a relative misalignment of the sensitive axes of the sensors also leads to a reduction in the value of $g^{(1)}_{AB}(0)$ (see Appendix\,\ref{app:AngleDep}). 
In order to distinguish a correlated signal from uncorrelated noise,  $g^{(1)}_{AB}(0)$ can be compared with $g^{(1)}_{AB}(\tau \gg \tau_\varphi)$. 




\medskip

\noindent \emph{Accessible UBDM parameter space.~---~}Existing sensor networks have sufficient sensitivity to probe experimentally unexplored parameter space describing UBDM by searching for correlated stochastic fluctuations using intensity interferometry.

\begin{figure}[t]
\center
\includegraphics[width=\columnwidth]{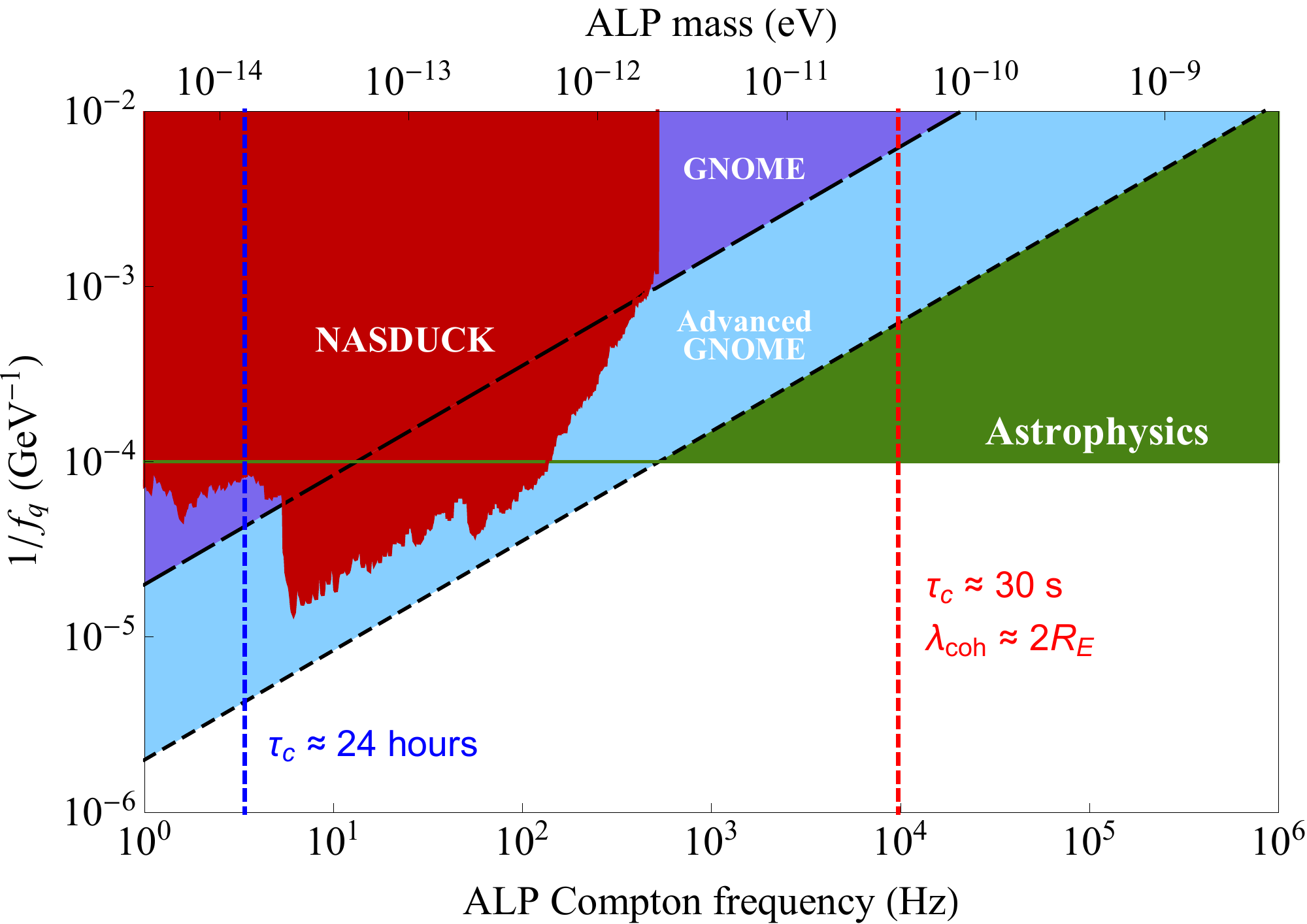}
\caption{\small{Estimated parameter space describing ALP dark matter that can be probed by GNOME (dashed line, purple shaded region) and Advanced GNOME (dotted line, light blue shaded region) measuring for $\approx 100$~days using $N_m = 10$ magnetometers \cite{Pus13,afach2018characterization,afach2021search,afach2023can}. GNOME and Advanced GNOME are sensitive to the interaction of the ALP field with proton spins described by Eq.\,\eqref{eq:quadratic-Hamiltonian}; $f_q$ parameterizes the ALP-nucleon coupling strength. The vertical dashed red line marks the Compton frequency and mass for which the ALP coherence length equals the Earth's diameter. The vertical dashed blue line marks the Compton frequency and mass for which $\tau_c \approx 24~{\rm hours}$. The dark red shaded region shows constraints from the Noble and Alkali Spin Detectors for Ultra-light Coherent darK matter (NASDUCK) experiment \cite{bloch2022new}. The dark green shaded area represents astrophysical bounds on spin-dependent ALP interactions with nucleons \cite{Oli08,chang2018supernova}. Note, however, that there are theoretical scenarios where these astrophysical bounds can be circumvented \cite{derocco2020exploring}.}}
\label{Fig:sensitivity-GNOME}
\end{figure}
\begin{figure}[t]
\center
\includegraphics[width=\columnwidth]{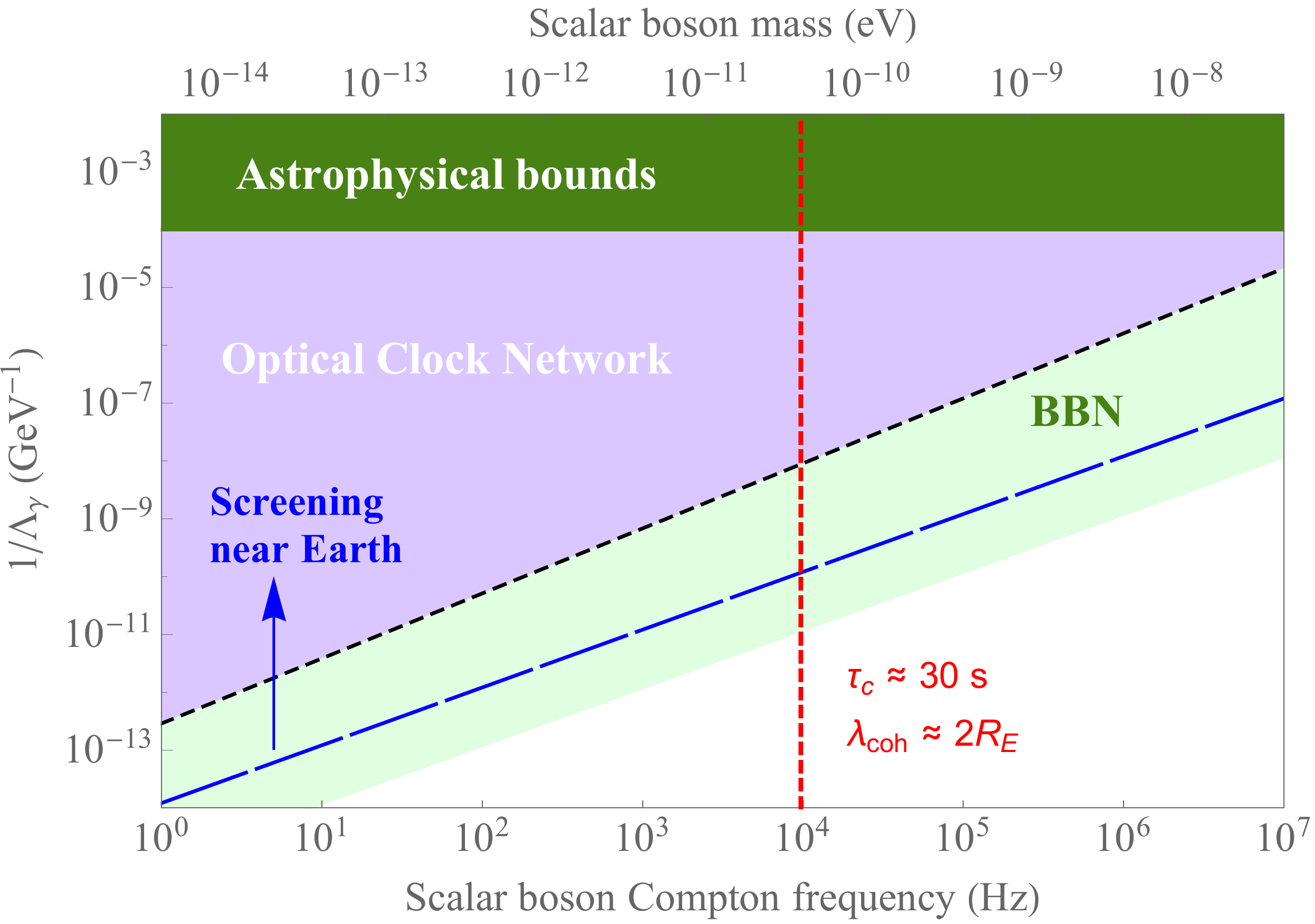}
\caption{\small{Estimated parameter space describing UBDM fields that can be probed by an optical clock network such as those described in Refs.\,\cite{collaboration2021frequency,lisdat2016clock} (dotted line, light purple shaded region) in $\approx 100$~days of searching for correlated stochastic fluctuations using $N_c = 10$ clocks, not accounting for (anti)screening from back-action \cite{hees2018violation,stadnik2020new}, which can play a significant role near Earth's surface above the long-dashed blue line as indicated by the blue arrow. Clocks are sensitive to the interactions described by Eq.\,\eqref{eq:scalar-Lagrangian}; $\Lambda_\gamma$ parameterizes the strength of the coupling of the ultralight bosons to photons. The vertical dashed red line marks the Compton frequency and mass for which the ultralight boson's coherence length equals the Earth's diameter. The dark green shaded area represents astrophysical bounds on such quadratic scalar interactions between ultralight bosons and photons from stellar cooling and observations of supernova 1987a \cite{Oli08,chang2018supernova}; the light green shaded region represents bounds from Big Bang nucleosynthesis (BBN) \cite{stadnik2015can}.}}
\label{Fig:sensitivity-clocks}
\end{figure}

Atomic magnetometers can search for ALP fields by detecting Zeeman shifts caused by the interaction described in Eq.\,\eqref{eq:quadratic-Hamiltonian}. In analogy with the Zeeman Hamiltonian, the gradient of the square of the ALP field acts as a ``pseudo-magnetic field'' $\bs{\mc{B}}_q$ given by
\begin{align}
    \bs{\mc{B}}_q \approx \mp\frac{2 \hbar^2 c^2}{g_F\mu_B f_q^2} \bs{\nabla} \varphi^2(\bs{r},t)~,
    \label{eq:pseudo-magnetic-field}
\end{align}
where $g_F$ is the Land\'e factor, $\mu_B$ is the Bohr magneton (or, for nuclear-spin-based magnetometers, the nuclear magneton). The projection of the near-dc component of $\bs{\mc{B}}_q$ along the sensitive axis of a magnetometer defined by the unit vector $\hat{\bs{m}}$ can be estimated in a manner similar to that discussed in, for example, Ref.\,\cite{gramolin2021spectral}, by evaluating the sum in Eq.\,\eqref{eq:gradient-field-sqr}, yielding a characteristic magnitude (with an average value of zero) of 
\begin{align}
    \hat{\bs{m}} \cdot \bs{\mc{B}}_q \sim \frac{\hbar^3 \rho\ts{dm} v_0}{g_F \mu_B m_\varphi f_q^2}~.
    \label{eq:ALP-intensity-gradient}
\end{align}
In the derivation of the above equation we assumed that frequency and wave-vector are uncorrelated. The accuracy of this approximation is discussed in Appendix~\ref{app:labframe}. Nonetheless, Eq.\,\eqref{eq:ALP-intensity-gradient} is suitable for a rough estimate of the sensitivity of a magnetometer network to such UBDM, given that $|\bs{v}\ts{lab}| \approx v_0$.

The sensitivity of a sensor network depending on the number of sensors ($N_m$), the UBDM field coherence time ($\tau_\varphi$), and total acquisition time ($T$) is discussed in Appendix~\ref{app:search}. Combining Eqs.\,\eqref{eq:ALP-intensity-gradient} and \eqref{eq:min-field}, an estimate for the UBDM coupling constant to which a magnetometer network would be sensitive is given by
\begin{align}
    f_q^2 \lesssim \frac{\hbar^3 \rho\ts{dm} v_0 }{g_F\mu_B m_\varphi \delta B} \prn{\tau_\varphi T}^{1/4} \sqrt{N_m}~.
    \label{eq:f_q-sens-mags}
\end{align}
Figure~\ref{Fig:sensitivity-GNOME} shows sensitivity estimates for the Global Network of Optical Magnetometers for Exotic physics searches (GNOME) \cite{Pus13,afach2018characterization,afach2021search,afach2023can} based on alkali vapor magnetometers with $\delta B \approx 100~{\rm fT/\sqrt{Hz}}$ and the Advanced GNOME network based on noble gas comagnetometers with $\delta B \approx 1~{\rm fT/\sqrt{Hz}}$, assuming $T=100$~days and $N_m = 10$. Note that for $\tau_c \gg 24$\,hours, the signal amplitude is partially modulated at the frequency of Earth's rotation since the signals are $\propto \bs{\hat{m}}\cdot\bs{\mc{B}}_q$ and $\hat{\bs{m}}$ rotates with the Earth while $\bs{\mc{B}}_q$ does not, which can in principle enable the detection of UBDM with coherence times much longer than a day. Notable is the extent to which GNOME and Advanced GNOME can probe UBDM with Compton frequencies far beyond the nominal sensor bandwidths.

Optical atomic clocks are an example of a sensor that can search for scalar fields through the apparent variation of fundamental constants as described in Eqs.\,\eqref{eq:scalar-mass-variation} and \eqref{eq:scalar-alpha-variation}, due to, for example, the variation of the fine-structure constant $\alpha$ and relativistic effects (see Ref.\,\cite{safronova2018search} and references therein). For example, the fractional frequency variation in an atomic clock due to variation of the fine-structure constant $\alpha$ is given by
\begin{align}
    \frac{\delta \nu(t)}{\nu} = \kappa_\alpha \frac{\delta \alpha(t)}{\alpha}~,
    \label{eq:frac-freq-variation-alpha-1}
\end{align}
where $\nu$ is the clock frequency and $\kappa_\alpha$ is a dimensionless sensitivity coefficient that depends on the type of clock: $\kappa_\alpha \approx 2$ for most current optical atomic clocks, but note that there are exceptions, such as the proposed clock with $\kappa_\alpha \approx -15$ described in Ref.\,\cite{Safronova2018Apr}, and the possibility of future clocks based on highly charged ions \cite{kozlov2018highly} or a Th nuclear transition \cite{peik2021nuclear} that could have orders of magnitude larger values of $\kappa_\alpha$. 

It is important to note that the sensitivity of intensity interferometry can be significantly impacted by back-action of the surrounding matter density on the scalar field as pointed out in Refs.\,\cite{stadnik2020new,hees2018violation} and also discussed in Refs.\,\cite{Oli08,hinterbichler2010screening,jaffe2017testing}. The accessible range of parameter space for current optical clocks near the surface of the Earth is well within the regime where such back-action effects are significant (above the long-dashed blue line in Fig.\,\ref{Fig:sensitivity-clocks}, see Appendix\,\ref{app:back-action}). However, for the range of boson masses and coupling constants considered in the present work, it turns out that for a space-based network of sensors \cite{Schkolnik2022Apr} the screening effects can be largely neglected \cite{stadnik2020new}, and so for simplicity we consider such a space-based network in our sensitivity estimates.

Assuming the effect described by Eq.\,\eqref{eq:scalar-alpha-variation} and a scalar field that makes up the entirety of the dark matter density, the amplitude of the fractional frequency variation is given by
\begin{align}
    \frac{\delta \nu}{\nu} \approx \kappa_\alpha \frac{2 \hbar^3 \rho\ts{dm}}{\Lambda_\gamma^2 m_\varphi^2 c}~.
    \label{eq:frac-freq-variation-alpha-2}
\end{align}
Optical clock networks, with $N_c$ independent clocks, can achieve a fractional frequency uncertainty~\cite{collaboration2021frequency,lisdat2016clock}
\begin{align}
    \frac{\delta \nu}{\nu} \approx \frac{3 \times 10^{-16}}{(\tau_\varphi T)^{1/4}\sqrt{N_c}}~,
    \label{eq:frac-freq-variation-clock-sens}
\end{align}
which translates to the sensitivity to the quadratic scalar coupling constant shown in Fig.\,\ref{Fig:sensitivity-clocks}. Appendix~\ref{app:SensiticityDifference} offers a heuristic argument for the significant  sensitivity difference between atomic clock and magnetometer networks to the respective coupling parameters $\Lambda_\gamma$ and $f_q$.



\medskip

\noindent\emph{Conclusion.~---~}In summary, we propose a new method to search for UBDM by using intensity interferometry with sensor networks. We show that when the sensors measure signals quadratic in the UBDM field, there is a near-dc component of the signal that enables finite-bandwidth sensors to search for UBDM with Compton frequencies many orders of magnitude larger than possible if a traditional search for signals oscillating at the Compton frequency is carried out. Here we have focused on quadratic UBDM interactions and sensors with a linear response; however, the results are also valid for linear interactions and sensors that respond quadratically to the field (square-law detectors). The method of intensity interferometry is intrinsically broadband, with the potential to search for UBDM with particle masses ranging over many orders of magnitude without having to probe individual narrow frequency bands. UBDM searches with intensity interferometry using existing sensor networks can probe unexplored parameter space.


\vspace{12pt}
The authors are sincerely grateful to Grzegorz Lukasiewicz, Jason Stalnaker, Ibrahim Sulai, Maxim Pospelov, Arran Phipps, Ben Buchler, Ron Folman, and Menachem Givon for enlightening discussions. The authors also thank Tatum Wilson, Rayshaun Preston, Christopher Verga, and Mario Duenas for early work on simulations. This article is based in part upon work from COST Action COSMIC WISPers CA21106, supported by COST (European Cooperation in Science and Technology). This work was supported by the U.S. National Science Foundation under grant PHYS-2110388, by the Cluster of Excellence ``Precision Physics, Fundamental Interactions, and Structure of Matter'' (PRISMA+ EXC 2118/1) funded by the German Research Foundation (DFG) within the German Excellence Strategy (Project ID 39083149), by the German Federal Ministry of Education and Research (BMBF) within the Quantumtechnologien program (Grant No. 13N15064), by the European Research Council (ERC) under the European Union Horizon 2020 research and innovation program (project Dark-OST, grant agreement No 695405), and by the DFG Project ID 423116110.
The authors at Boston University acknowledge support from the Simons Foundation Grant No. 641332, the National Science Foundation CAREER Award No. PHY-2145162, the John Templeton Foundation Grant No. 60049570, and the U.S. Department of Energy, Office of High Energy Physics program under the QuantISED program, FWP 100495. The work of YVS was supported by the Australian Research Council under the Discovery Early Career Researcher Award DE210101593. The work of SP was supported by the National Science Centre of Poland within the grant No. 2020/39/B/ST2/01524.

\bibliography{stochastic-UBDM-search-bib.bib}

\begin{thebibliography}{100}
\bibitem{TruckNote}
Note that the spatial dependence for coherence lengths smaller than the Earth's
  diameter could also be exploited to look for UBDM, for which portable sensors
  could be employed as in Refs.\,\cite{grotti2018geodesy,takamoto2020test}.
\end{thebibliography}
\clearpage 
\appendix
\section{The effect of velocity of the laboratory frame}
\label{app:labframe}


The velocity offset due to the movement of the laboratory frame, here equivalent to the galactic velocity of Earth $\bs{v}_\ts{lab} = v_\ts{lab} \bs{\hat{z}}$, breaks the isotropy of the field. However, this anisotropy is not directly observed when measuring the average intensity because the quadratic interactions depend on the difference of wave-vectors and frequencies as can be seen in Eqs.\,\eqref{eq:field-squared}\,and\,\eqref{eq:gradient-field-sqr}. Yet, the coherence time of the field is different for directions parallel and perpendicular to $\bs{v}_\ts{lab}$ and depends on its magnitude.

\begin{figure}[ht]
	\center
	\includegraphics[width=\columnwidth]{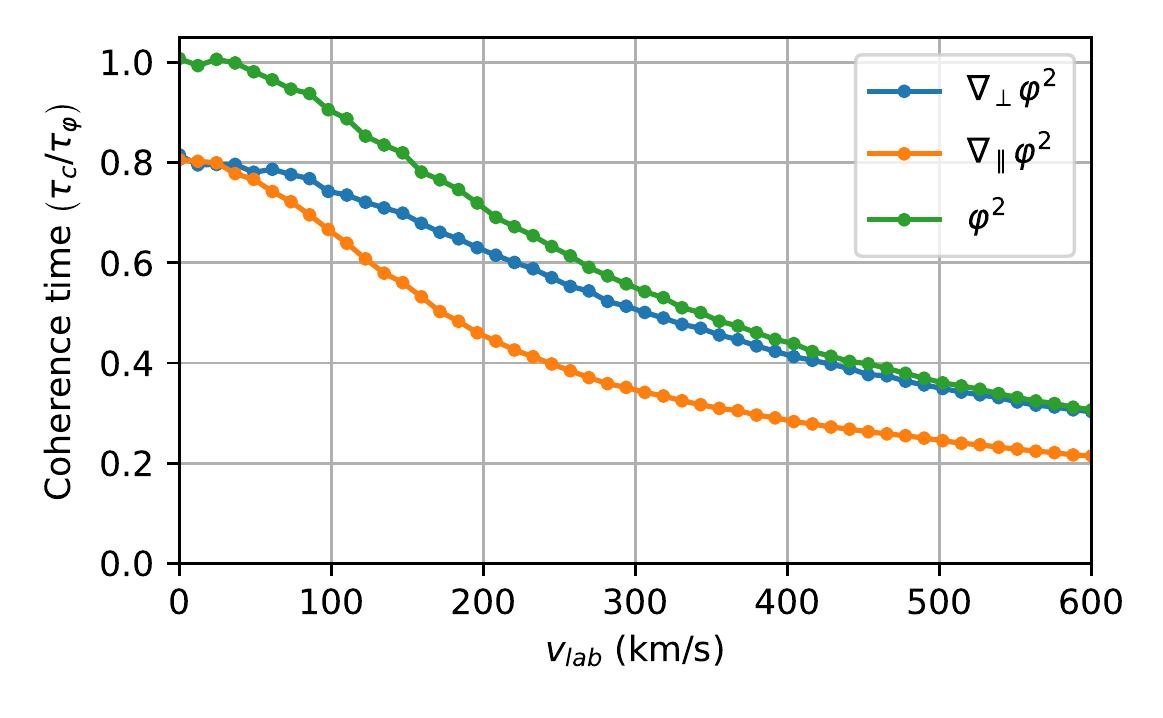}
	\caption{\small{ Coherence time, as defined in Eq.\,\eqref{eq:coherence_time_def}, as a function of $\bs{v}_\ts{lab}$ for the different cases of the UBDM field $\varphi$ studied.}}
	\label{Fig:tcVSvab}
\end{figure}

To understand the influence of $\bs{v}_\ts{lab}$ on the coherence time it is instructive to consider the limiting cases: $\bs{v}_\ts{lab}=0$ and $| \bs{v}_\ts{lab} | \gg v_0$ . Decomposing $\bs{k}$ into parallel and perpendicular components to $\bs{v}_\ts{lab}$ yields the following relationships:
\begin{equation}
	\bs{k}^\parallel_{nm} = \frac{ m_\varphi \bs{v}^\parallel_{nm} }{\hbar},~\text{and}~\bs{k}^\perp_{nm} = \frac{ m_\varphi \bs{v}^\perp_{nm} }{\hbar}\,.
	\label{eq:k_nm}
\end{equation}
The velocity difference  $\bs{v}_{nm} = \bs{v}_{n}-\bs{v}_{m}$ and the superscripts $\parallel$ and $\perp$ indicate whether the component is parallel or perpendicular to $\bs{v}_\ts{lab}$. One can also write the difference of frequencies in terms of the galactic-rest-frame velocity components as
\begin{equation}
	\begin{split}
		\frac{2c^2\omega_{nm}}{\omega_c} =~&(\bs{v}^\parallel_{n} + \bs{v}^\parallel_{m})\cdot \bs{v}^\parallel_{nm} + (\bs{v}^\perp_{n} + \bs{v}^\perp_{m})\cdot\bs{v}^\perp_{nm} \\
		&+ 2\bs{v}_\ts{lab}\cdot\bs{v}^\parallel_{nm}.
		\label{eq:omega_nm}
	\end{split}
\end{equation}

For $\bs{v}_\ts{lab}\ll \bs{v}_\ts{0}$ there is almost no anisotropy and therefore parallel and perpendicular components of $\nabla\varphi^2(\bs{r},t)$ have the same coherence time. However, as can be seen in Eqs.\,\eqref{eq:k_nm} and \eqref{eq:omega_nm}, $\omega_{nm}$ and $\bs{k}_{nm}$ have different dependence on $\bs{v}_{nm}$. For $\bs{k}_\ts{nm}$ the relationship is linear, while for $\omega_\ts{nm}$ it is non-linear and depends on the product of the sum and the difference of velocities. This implies a nonlinear correlation between $\omega_{nm}$ and $\bs{k}_{nm}$ as can be seen in Fig.\,\ref{Fig:tcVSvab}. Larger $\bs{v}_{nm}$ correlates with larger $\omega_{nm}$ and $\bs{k}_{nm}$. Since the gradient coupling is weighted by $\bs{k}_{nm}$ [see Eq.\,\eqref{eq:gradient-field-sqr}], the larger $\omega_{nm}$ have more weight in the sum. This broadens the frequency spectrum and shortens the coherence time of the gradient with respect to the field squared as can be seen in Fig.\,\ref{Fig:tcVSvab}. 

For large $v_\ts{lab}$ the behavior of $\nabla_\perp\varphi^2$ resembles that of $\varphi^2$ (see Fig.\,\ref{Fig:tcVSvab}) because the perpendicular terms of the velocity $\bs{v}^\perp_{nm} $ contribute little to $\omega_{nm}$. Therefore, $\bs{k}^\perp_{nm}$ and $\omega_{nm}$ are effectively independent and Eq.\,\eqref{eq:gradient-field-sqr} resembles Eq.\,\eqref{eq:field-squared} if multiplied by an appropriate scaling factor. 

\begin{figure}[ht]
	\center
	\includegraphics[width=\columnwidth]{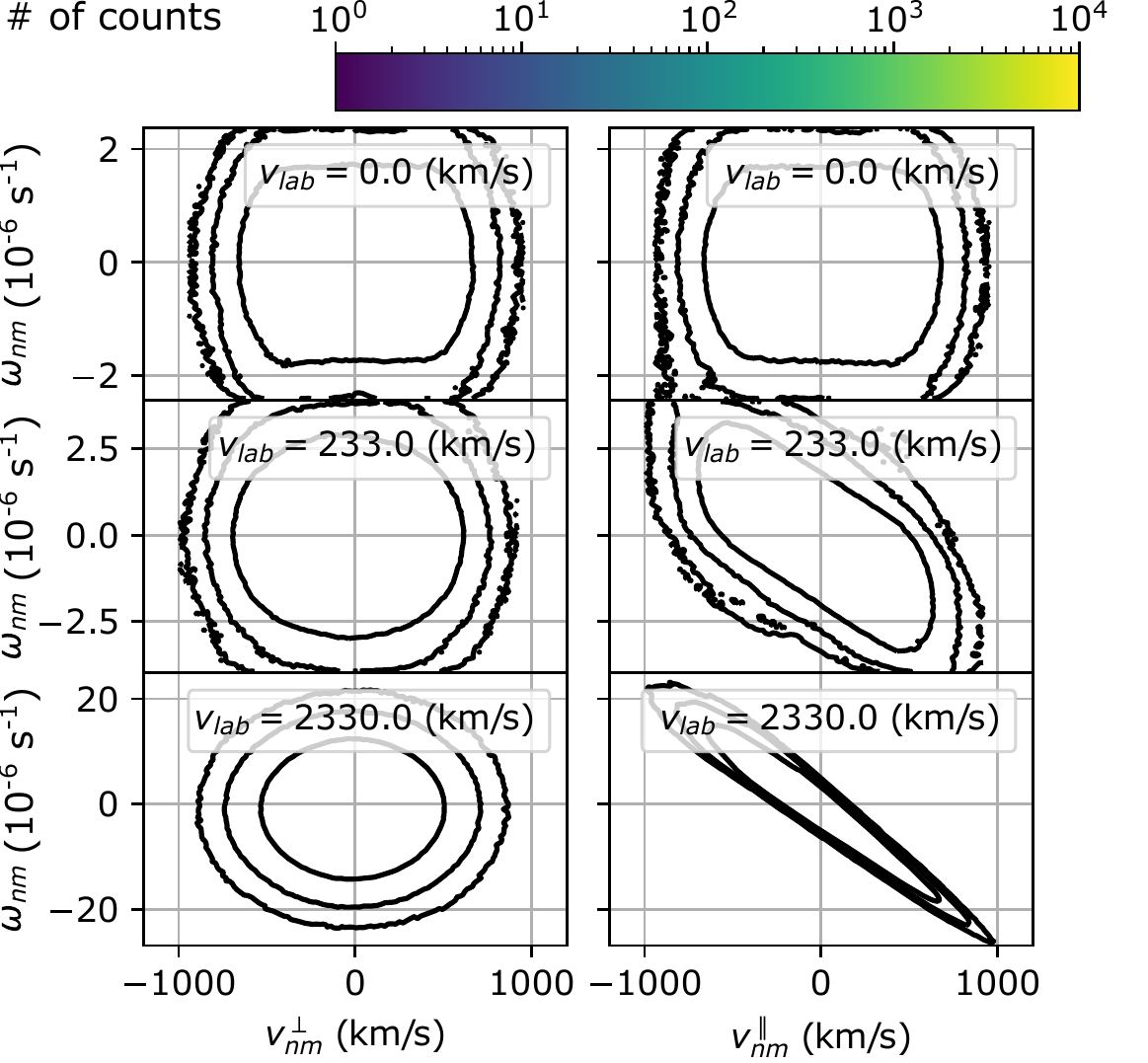}
	\caption{\small{ Visualization of the dependence of $\omega_{nm}$ on $\bs{v}_{nm}^\perp$ and $\bs{v}_{nm}^\parallel$. The Monte Carlo simulation was performed drawing $10^5$ velocities from the distribution in Eq.~\eqref{eq:f_lab_velocity}. The number of instances in which $\bs{v}^\perp_{nm} $ or $\bs{v}^\parallel_{nm}$ results in a $\omega_{nm}$ within a 20 km/s $\times 4.6 \times 10^{-7}$ s$^{-1}$ bin is represented in color. The black contour-lines indicate when the number of counts is $10^1$, $10^2$ or $10^3$ and serve as a guide to illustrate the shape of the distributions. }}
	\label{fig:correlation}
\end{figure}

In contrast, the coherence time of $\nabla_\parallel\varphi^2$ reduces faster than that of $\varphi^2$ as can be seen in Fig.\,\ref{fig:correlation}. The reason for this is that for large $v_\ts{lab}$, $\omega_{nm}$ becomes approximately $\propto \bs{k}^\parallel_{nm}$ and this proportionality scales with $v_\ts{lab}$. This increases the weight in the sum of terms with large $\omega_{nm}$ [in Eq.\,\eqref{eq:gradient-field-sqr}], leading to a broader power spectrum and consequently, a shorter coherence time. 

In the derivation of Eq.~\eqref{eq:ALP-intensity-gradient} in the main text we assumed that the average spread of values of $\hat{\bs{m}}\cdot\bs{k}_{nm}$ is $\approx m_\varphi v_0/\hbar$. This estimate implicitly assumes that $\omega_{nm}$ and $\bs{k}_{nm}$ are uncorrelated, and thus the accuracy of this approximation varies with $\bs{v}\ts{lab}$. Nonetheless, Eq.\,\eqref{eq:ALP-intensity-gradient} is suitable for a rough estimate of the sensitivity of a magnetometer network to such UBDM, given that $|\bs{v}\ts{lab}| \approx v_0$.

\section{Search with correlated sensor networks}\label{app:search}

The stochastic properties of $\varphi^2_s$ and $\bs{\nabla}\varphi^2_s$ can be used to search for UBDM. However, magnetometers and optical clocks are subjected to systematic noise sources in the laboratory, which feature some degree of temporal self-coherence similar to the UBDM field. For this reason, we focus our search on the cross-correlation of independent sensors, as their noise can be assumed to be mostly uncorrelated. 

Prosaic natural phenomena may also limit the sensitivity of a sensor network. Known natural phenomena that could generate noise with long-range correlations include, for example, time-dependent electromagnetic fields associated with resonances of the conducting Earth-ionosphere cavity \cite{fedderke2021earth} (such as the Schumann resonances \cite{schumann1952strahlungslosen}) and vibrational noise due to free oscillations of the Earth excited by large earthquakes \cite{hu2020network} (such as the ``breathing'' mode of the Earth at $\approx 800~{\rm \mu Hz}$ \cite{dziewonski1981preliminary,rosat2007geographical}). Auxiliary measurements with other instruments may be able to rule out such systematic effects. For example, GNOME uses unshielded magnetometers to monitor the magnetic environment near the shielded dark matter sensors to veto signals from anomalously large local magnetic field excursions \cite{afach2018characterization}. Additionally, as showed in Appendix~\ref{app:labframe}, the coherence time of the UBDM has a dependence on the measurement axis relative to $\bs{v}_{\text{lab}}$. The characteristic daily and annual modulation of the coherence time due to Earth motion can be used to confirm a UBDM signal. The possibility of the data containing non-dark-matter long-range correlated signals is not considered in this analysis. 

In order to illustrate a possible method that could be used to search for UBDM fields using intensity interferometry, let us consider a measured time-series $\tilde{\mc{S}}_{A,B}(t)$ lasting several $\tau_{\varphi}$ from two different sensors, $A$~and~$B$. These time-series have their mean subtracted, such that $\mc{S}_{A,B} = \tilde{\mc{S}}_{A,B} - \langle \tilde{\mc{S}}_{A,B} \rangle_t$. The measurement in each sensor would have a small UBDM-related component $\kappa s(t)$, and a noise term $\mc{N}(t)$, $\mc{S}_{A,B} = \kappa_{A,B} s_{A,B}(t) + \mc{N}_{A,B}(t)$. The factor $\kappa_{A,B}$ accounts for coupling of the sensor to the UBDM field, and $s_{A,B}(t)$ is either $\varphi^2_s$ (scalar interaction) or $\hat{\bs{m}} \cdot \bs{\nabla} \varphi^2_s$ (pseudoscalar gradient interaction).

For a particular time series the degree of first-order coherence of these signals can be calculated as
\begin{equation}
	g^{(1)}_{AB}(\tau) = \frac{\langle \mc{S}_A(t) \mc{S}_B(t+\tau) \rangle_t}{\sqrt{\langle \mc{S}_A^2 \rangle_t \langle \mc{S}_B^2 \rangle_t}}\,.
	\label{eq:g1appc}
\end{equation}
While the signals $s_A(t)$ and $s_B(t)$ are correlated, the corresponding noise contributions, $\mc{N}_A(t)$ and $\mc{N}_B(t)$, are not. As a consequence, one can suppress the cross-terms $\langle \mc{N}_A \mc{N}_B \rangle_t$, $\langle \mc{N}_A s_B \rangle_t$, and $\langle \mc{N}_B s_A \rangle_t$ with sufficient amount of averaging (determined by the signal-to-noise ratios of sensors $A$ and $B$). Then, Eq.\,\eqref{eq:g1appc} reduces to
\begin{equation}
	g^{(1)}_{AB}(\tau) = \kappa_A \kappa_B \frac{ \langle  s_A(t) s_B(t+\tau) \rangle_t}{\sqrt{\langle \mc{S}_A^2 \rangle_t \langle \mc{S}_B^2 \rangle_t}}\,.
\end{equation}

Strictly speaking, there will be a difference between $s_A(t)$ and $s_B(t)$ because of the spatial dependence of $\varphi^2$  (or $\bs{\nabla} \varphi^2$) that arises from the $\bs{k}\cdot \bs{r}$ term in Eq.\,\eqref{eq:single-boson-field}. However, for the considered UBDM mass range, the coherence length is much larger than the distance between the sensors. Therefore, $\varphi^2_s$ (or $\bs{\nabla} \varphi^2_s$) will approximately be the same for both sensors, $s_{A,B}(t) \approx s(t)$; and so $g^{(1)}(\tau)$ is proportional to the autocorrelation of the slow-varying component of the UBDM field squared~(or its gradient):
\begin{equation}
	\label{eq:g_tau_simple}
	g^{(1)}_{AB}(\tau) \approx \frac{\kappa_A \kappa_B}{\sqrt{\langle \mc{S}_A^2 \rangle_t \langle \mc{S}_B^2 \rangle_t}}\langle s(t) s(t+\tau) \rangle_t.
\end{equation}

As an example, evaluating Eq.\,\eqref{eq:g_tau_simple} at $\tau=0$ yields
\begin{equation}
	g^{(1)}_{AB}(0) \approx \frac{\kappa_A \kappa_B}{\sqrt{\langle \mc{S}_A^2 \rangle_t \langle \mc{S}_B^2 \rangle_t}}\langle s^2 \rangle_t .
	\label{eq:g1ofzero}
\end{equation}
The quantity $g^{(1)}(0)$ can be used as an estimator for the slow-varying component (of the gradient) of the field squared; it can be used to determine whether or not a UBDM signal is present in the data. For the scalar interaction, $\langle s^2 \rangle_t = \langle \varphi^2_s \varphi^2_s  \rangle_t$ and it can be shown  that for the gradient interaction $\langle s^2 \rangle_t = \prn{\cos \vartheta_{AB}/3} \langle \bs{\nabla}\varphi^2_s \cdot \bs{\nabla}\varphi^2_s  \rangle_t$, where $\vartheta_{AB}$ is the angle between the sensitive axes of the two sensors, $\bs{\hat{m}}_A$ and $\bs{\hat{m}}_B$ (see Appendix~\ref{app:AngleDep}). Assuming Gaussian noise with standard deviation $\sigma$, $\mc{N}\sim \mc{G}(0,\sigma^2)$, that dominates over the UBDM-signal component, $ \kappa s \ll \sigma$, Eq.\,\eqref{eq:g1ofzero} can be simplified to:
\begin{equation}
	g^{(1)}_{AB}(0) \approx \frac{\kappa_A \kappa_B}{\sigma_A \sigma_B}\langle s^2 \rangle_t.
	\label{eq:g0approx}
\end{equation}
The presence of a common correlated UBDM signal between two sensors will lead to a nonzero $g^{(1)}(0)$. This method is inherently broadband, as a large range of UBDM masses could lead to $|g^{(1)}(0)|>0$. 

In order to estimate the sensitivity to a nonzero $g^{(1)}(0)$, we assume $N_m$ identical sensors having the same directional sensitivity and dominated by Gaussian noise with the same variance (of course, this is not the typical case for real networks, these feature different directional sensitivities --- see Appendix~\ref{app:AngleDep} --- and are affected by $1/f$ noise).
In the cases of interest, the coherence length of the UBDM is much larger than the spacing between the sensors, so the UBDM signal will be identical in all sensors. 
To understand the scaling of the sensitivity with $N_m$, $\tau_\varphi$, and total acquisition time $T$, suppose that we divide the network into two distinct groups each with $N_m/2$ sensors and average the data in each group. 
Furthermore, suppose that the time-series data are binned in $\tau_\varphi$-long segments and time-averaged in each bin. 
Based on this approach, in the absence of a correlated UBDM signal, each group of sensors will exhibit a bin-to-bin variance in the measured magnetic field of approximately $\delta B / \sqrt{\tau_\varphi N_m/2} \sim \delta B / \sqrt{\tau_\varphi N_m}$, where $\delta B$ is the sensitivity of a single magnetometer in units of magnetic field strength times the square root of time. For a total acquisition time $T$, there will be $N_b = T/\tau_\varphi$ bins.
The expected residual from noise in the cross-correlation between the two groups provides an estimate of the network resolution, which in turn gives the minimum detectable pseudo-magnetic field squared 
\begin{align}
    \abrk{ \mc{B}_q({\rm{min}})^2 }_t \approx \frac{\delta B^2}{N_m \tau_\varphi}\frac{1}{\sqrt{N_b}} \approx \frac{\delta B^2}{N_m \sqrt{\tau_\varphi T}}~.
    \label{eq:min-field}
\end{align}
An in-depth discussion of the $T$ and $N_m$ scaling for a similar case can be found in Ref.~\cite{derevianko2018detecting}.

This scheme does not rely on the boson having a Compton wave resonant with the experimental set-up. Such resonant searches require the interrogation of numerous narrow frequency bands which is a time-consuming process \cite{Asz01,Asz10,budker2014proposal,Sik14,chaudhuri2015radio,phipps2020exclusion,manley2020searching,crescini2020axion,braine2020extended,backes2021quantum,Ayb21CASPErE}.
In the case of detection in a search for correlated stochastic fluctuations, the mass range of the UBDM particle could be narrowed by analyzing the coherence time, as illustrated in Fig.~\ref{Fig:massdep}.
\section{Angle dependence of the correlation between magnetometers}\label{app:AngleDep}

We are interested in understanding the coherence between two signals, $s_A$ and $s_B$, that are the result of the gradient coupling. In particular, for $j\in\{A,B\}$, let $\boldsymbol{\hat{m}_j}$ be the sensitive axis of magnetometer $j$, so $s_j = \boldsymbol{\hat{m}_j} \cdot \boldsymbol{\nabla} \varphi^2_s$. The calculation of the coherence requires one to understand $\left< s_A s_B \right>_t$, but each $s_A$ and $s_B$ depends on the relative angle between the respective sensitive axis and the field gradient. These angles are not known and change over time as the field evolves.

Note that $\boldsymbol{\nabla} \varphi^2_s$ is a sum of terms $\propto \bs{v}_{nm}$.  Additionally, since the field gradient experiences the same bias velocity at the location of both sensors, the relative velocities $\bs{v}_{nm}$ have no bias. The angle dependence between the field gradient and the sensitive axes can be accounted for under some basic assumptions. In particular, assume that $\boldsymbol{v}_{nm}$ is ergodic in direction --- that is, over time, $\boldsymbol{\hat{v}}_{nm}$ will take all values over the sphere $S^2$.
Ergodicity implies that $\left< s_A s_B\right>_t = \left< s_A s_B\right>_{S^2}$. 
As another simplifying assumption, let the relative speed $v_{nm}$ be independent of the direction $\boldsymbol{\hat{v}}_{nm}$ so 
\[
\left< (\boldsymbol{\hat{m}_A}\cdot \boldsymbol{v}_{nm})(\boldsymbol{\hat{m}_B}\cdot \boldsymbol{v}_{nm}) \right>_t = \bar{v}^2 \left< (\boldsymbol{\hat{m}_A}\cdot \boldsymbol{\hat{v}}_{nm})(\boldsymbol{\hat{m}_B}\cdot \boldsymbol{\hat{v}}_{nm}) \right>_t\,,
\]
for $\bar{v}^2 = \left< \lVert \boldsymbol{v}_{nm} \rVert^2 \right>_t$\,.

The correlation between magnetometers can be calculated explicitly with the above assumptions. This will still depend on the relative angle $\vartheta_{AB}$ between the sensitive axes $\boldsymbol{\hat{m}_A}$ and $\boldsymbol{\hat{m}_B}$. Let us define the axes such that $\boldsymbol{\hat{m}_B}$ points along the $z$-axis and $\boldsymbol{\hat{m}_A}$ lies in the $xz$-plane in the $+x$-direction. The gradient direction $\boldsymbol{\hat{v}}$ can be arbitrary, described by the azimuthal angle $\phi$ and polar angle $\theta$ in this coordinate system. Thus
\begin{align*}
	\boldsymbol{\hat{m}_A} &= \left( \sin\vartheta_{AB}, 0, \cos\vartheta_{AB} \right)\ , \\
	\boldsymbol{\hat{m}_B} &= (0, 0, 1)\ , \\
	\boldsymbol{\hat{v}}_{nm} &= \left( \cos\phi \sin\theta, \sin\phi \sin\theta, \cos\theta \right)\ .
\end{align*}
The following calculation can be made explicitly:
\begin{align}
	\left<s_As_B\right>_t &\approx \left<s_As_B\right>_{S^2} \nonumber \\
	&= \frac{\bar{v}^2}{4\pi}\int_0^{2\pi}d\phi \int_0^\pi d\theta ( \sin\vartheta_{AB} \cos\phi\sin\theta \nonumber \\
	& \qquad +\cos\vartheta_{AB} \cos\theta )\cos\theta\sin\theta \nonumber \\
	&= \frac{\bar{v}^2}{3} \cos\vartheta_{AB}\ ,
	\label{eq:angledep}
\end{align}
where the $\sin\vartheta_{AB}$ term averages out over the sphere over both the azimuthal and polar directions. 

\begin{figure}
	\center
	\includegraphics[width=\columnwidth]{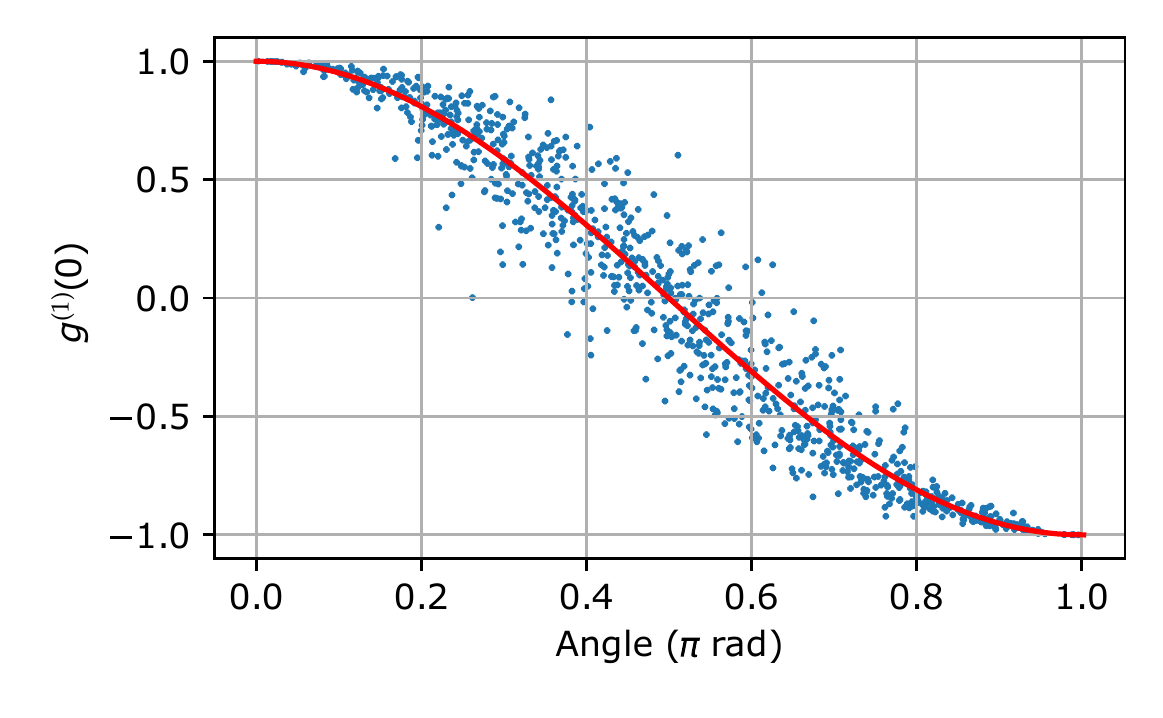}
	\caption{\small{$g^{(1)}(0)$ of $\hat{\bs{m}}_A \cdot \bs{\nabla} \varphi_s^2$ and $\hat{\bs{m}}_B \cdot \bs{\nabla} \varphi_s^2$ as a function of the relative angle between $\hat{\bs{m}}_A$ and $\hat{\bs{m}}_B$, $\vartheta_{AB}$. No noise was added to the generated fields. The blue dots are the results of independent simulations where the sensitive axes are randomly generated on the sphere. The red curve shows the $\cos(\vartheta_{AB}$) as predicted in Eq.\,\eqref{eq:angledep}.}}
	\label{Fig:angledep}
\end{figure}

The results of simulations shown in Fig.\,\ref{Fig:angledep} illustrate this dependence. An interesting feature is that the dispersion in $g^{(1)}(0)$ also depends on the relative angle between the magnetometers. This dispersion in $g^{(1)}(0)$ is not due to noise, as noise was not added to the simulated magnetometer data, but rather results from the UBDM fields themselves. If the sensitive axes of the sensors are not parallel,  dispersion in $g^{(1)}(0)$ is introduced by the uncorrelated components of the UBDM field measured; because the orthogonal components of $\bs{\nabla} \varphi_s^2$ (e.g., along $\hat{x},\hat{y},\hat{z}$) are uncorrelated, measuring different orientations produces a spread in $g^{(1)}(0)$. This characteristic feature may be a useful signature that could be used for validating a UBDM signal.

\section{Analytic solution for \texorpdfstring{$g^{(1)}(\tau)$}{g1(t)}}
\label{app:analytic}

In this Appendix, an analytic expression for the degree of coherence is derived. Let us consider a signal given by the sum of sinusoidal functions,
\begin{equation}\label{eq:sinSumSig}
	S(t) = \sum_n a_n \cos\left( \omega_n t + \phi_n \right)\,.
\end{equation}
For simplicity, we consider the case in which the average signal is zero, $\omega_n\neq0$, and the frequencies are all different, $\omega_n\neq\omega_m$ for $m\neq n$. Considering the two-point correlation function defined as,
\begin{equation}\label{eq:2pt_correlation}
	C_2(\tau) = \left< S(t)S(t+\tau) \right>_t\,,
\end{equation}
then $g^{(1)}(\tau) = C_2(\tau)/C_2(0)$. 

When averaging over the product of the two signal functions, the oscillating components average out. As a result, the only components that contribute to the coherence are from products of the same sinusoidal terms. The resulting two-point correlation function is then
\footnote{The relevant terms are of the form $C = \cos\left(x\right) \cos\left(x+y\right)$, for $x=\omega_n t + \phi_n$ and $y=\omega_n \tau$. Expanding 
\begin{align*}
C&=\cos\left(x\right) \left[ \cos\left(x\right)\cos\left(y\right) - \sin\left(x\right)\sin\left(y\right) \right] \\
&= \frac{1}{2}\left[ 1+\cos\left(2x\right) \right]\cos\left(y\right) - \frac{1}{2}\sin\left(2x\right)\sin\left(y\right)\,.
\end{align*}
Averaging over $t$, the sinusoidal terms with $x$ vanish, leaving $\frac{1}{2}\cos\left(y\right) = \frac{1}{2}\cos\left(\omega_n\tau\right)$.
}
\begin{equation}\label{eq:sinSum2ptCorr}
	C_2(\tau) = \frac{1}{2} \sum_n a_n^2 \cos\left( \omega_n\tau \right)\,.
\end{equation}

For this study, we consider the case in which there are $N\gg 1$ sinusoidal components in the signal. The frequencies $\omega_n$ and amplitudes $a_n$ are sampled from some probability distribution. Here, this distribution is derived from a distribution of particle velocities $P_v(\boldsymbol{v})d^3\boldsymbol{v}$ given by Eq.\,\eqref{eq:f_lab_velocity}. 

We approximate the sum in Eq.\,\eqref{eq:sinSum2ptCorr} as an integral in which $a_n^2$ is replaced with the differential $a^2(\boldsymbol{x})P(\boldsymbol{x}) d\boldsymbol{x}$,
\begin{equation}\label{eq:sinInt2ptCorr}
	C_2(\tau) = \frac{1}{2} \int d\boldsymbol{x}\,a^2(\boldsymbol{x})P(\boldsymbol{x}) \cos\left[\omega(\boldsymbol{x}) \tau \right]\,.
\end{equation}

Here, for both the $\varphi^2$- and $\bs{\nabla}\varphi^2$-coupling, the probability distribution function is parameterized by $\boldsymbol{x}=\{ \boldsymbol{k_1}, \boldsymbol{k_2} \}$, for $\boldsymbol{k_i}=(m_\varphi / \hbar)\boldsymbol{v_i}$ described by a Gaussian distribution of velocities offset by the laboratory reference frame. It is useful to consider the average $\boldsymbol{\bar{k}}=(\boldsymbol{k_1}+\boldsymbol{k_2})/{2}$ and difference $\boldsymbol{\Delta} = \boldsymbol{k_1} - \boldsymbol{k_2}$. In these variables, $d^3\boldsymbol{k_1}d^3\boldsymbol{k_2}=d^3\boldsymbol{\bar{k}}d^3\boldsymbol{\Delta}$ and the difference in frequencies $\omega_{12} = (\hbar/m_\varphi) \boldsymbol{\bar{k}}\cdot\boldsymbol{\Delta}$. For either coupling, the contribution from $\mathcal{O}(2\omega_c)$ frequency components and the constant offset will be neglected. Also, let $\sigma_k=(m_\varphi / \hbar)\sigma_v$ ($v_0^2=2\sigma_v^2$) describe the width of the $\boldsymbol{k}$-vector distribution and $\boldsymbol{k_\text{lab}} = ({m_\varphi}/{\hbar}) \boldsymbol{v_\text{lab}}$ describe the offset of the distribution due to the laboratory reference frame. 

\begin{figure}
    \centering
    \includegraphics[width=\columnwidth]{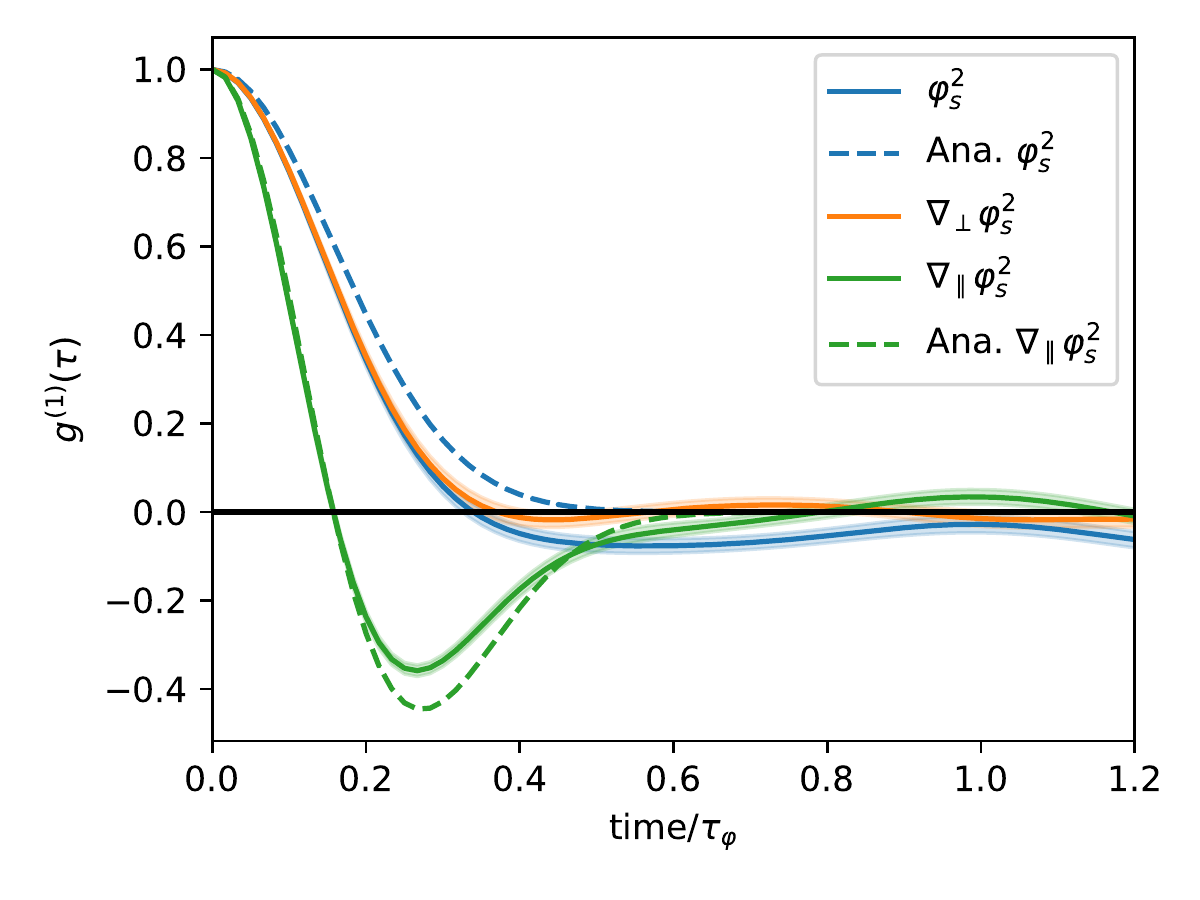}
    \caption{\small{Coherence of different signal types from numeric simulations (solid) compared to their analytic approximations (dashed, given by Eqs.\,\eqref{eq:g1_analytic_phi2}\,,\,\eqref{eq:g1_analytic_grad_phi2}). To better reflect the approximation conditions, the lab velocity is tripled, $v_\text{lab}=699$\,km/s (while $v_0$ is kept at $220$\,km/s). The analytic approximation for $\varphi_s^2$ is the same as for the gradient coupling for a $\boldsymbol{\hat{m}}$ perpendicular to the lab velocity $\nabla_\perp\varphi_s^2$.}}
    \label{fig:coherence_analytic}
\end{figure}

For a signal given by $\varphi_s^2$, the amplitude of a sinusoidal component is constant while the frequency is given by the difference $\omega(\boldsymbol{k}_1) - \omega(\boldsymbol{k}_2)$. The integral in Eq.\,\eqref{eq:sinInt2ptCorr} can be approximated for $k_\text{lab} \gg \sigma_k$. Under this approximation, $\boldsymbol{\bar{k}}\approx\boldsymbol{k_\text{lab}}$ (i.e., one can neglect the uncertainty in the average $\boldsymbol{\bar{k}}$), which reduces the six-dimensional integral to a three-dimensional integral over $\boldsymbol{\Delta}$. The result is the degree of first-order coherence (see Fig.\,\ref{fig:coherence_analytic})
\begin{equation}
	g^{(1)}(\tau) \approx \exp\left[ -\left( \frac{m_\varphi v_\text{lab} \sigma_v}{\hbar} \right)^2 \tau^2 \right]\,.
	\label{eq:g1_analytic_phi2}
\end{equation}
Note that we consider a lab-frame for which $v_\ts{lab}\approx v_k$, so this is only a rough approximation. Heuristically, one would expect that the coherence is overestimated, because this approximation neglects a component of the frequency dispersion. 

A signal given by coupling to the gradient $\nabla \varphi_s^2$ will have amplitudes of different sinusoidal components $\sim \boldsymbol{\hat{m}}\cdot\boldsymbol{\Delta}$. Here, we denote the angle between $\boldsymbol{\hat{m}}$ and $\boldsymbol{v_\text{lab}}$ with $\theta$. Using the approximation $k_\text{lab} \gg \sigma_k$ as before, the coherence is (see Fig.\,\ref{fig:coherence_analytic})
\begin{align}
	g^{(1)}(\tau) &\approx \frac{\sin^2(\theta) + 2\pi\left[ 1-2\left( \frac{m_\varphi}{\hbar} v_\text{lab} \sigma_v \tau \right)^2 \right]\cos^2(\theta) }{\sin^2(\theta) + 2\pi\cos^2(\theta) } \nonumber \\
	&\qquad\times\exp\left[ -\left( \frac{m_\varphi v_\text{lab} \sigma_v}{\hbar} \right)^2 \tau^2 \right]\,.
	\label{eq:g1_analytic_grad_phi2}
\end{align}
Observe that the coherence becomes negative for $\theta\neq \pi/2$ when
\begin{equation}
    \tau>\frac{\hbar}{m_\varphi v_\text{lab} \sigma_v}\sqrt{\frac{1}{2}+\frac{\tan^2(\theta)}{2\pi}}\,.
\end{equation}
This is a result of the fact that for such time shifts $S(t)$ and $S(t+\tau)$ are out-of-phase.

\section{Back-action effects}
\label{app:back-action}

As noted in the main text, a consideration particular to searches for ultralight scalar fields with quadratic couplings to SM particles and fields is the possible back-action of mass density on the scalar field, which can either reduce or enhance the field amplitude \cite{stadnik2020new,Oli08,hinterbichler2010screening,jaffe2017testing,hees2018violation}. This back-action arises due to the fact that the bare potential for the scalar field, namely $m_\varphi^2 c^2 \varphi^2/(2 \hbar^2)$, is modified by the presence of the Lagrangian in Eq.\,\eqref{eq:scalar-Lagrangian}. 

The quadratic terms in the interaction Lagrangian endow the scalar field with an effective mass $m\ts{eff}$ in the presence of matter; this effect is analogous to the screening of magnetic fields caused by the Meissner effect~(where the photon gains an effective mass inside a superconductor). Depending on the sign of the interaction, back-action can lead to either screening or antiscreening \cite{hees2018violation}.

For the interaction between $\varphi$ and the electromagnetic field leading to variation of $\alpha$ considered in Eq.~\eqref{eq:scalar-Lagrangian}, in the presence of matter the scalar field acquires an effective mass given by
\begin{align}
m^2_\ts{eff} \approx m_\varphi^2 \pm \frac{2 \hbar^3}{c} \frac{\rho_\gamma}{\Lambda_\gamma^2}~,
    \label{eq:effective-scalar-mass}
\end{align}
where $\rho_\gamma \approx -F_{\mu\nu}^2/4$ is the Coulomb energy density of a nonrelativistic nucleus averaged over the characteristic volume outside of which the UBDM field changes appreciably, which in our considered case is orders of magnitude larger than the size of a nucleus and thus depends on the average local matter density. Based on the numerical estimates of Ref.\,\cite{stadnik2020new}, for the Earth's interior $\rho_\gamma \approx 6 \times 10^{21}~{\rm GeV/cm^3}$ and for Earth's atmosphere $\rho_\gamma \approx 6 \times 10^{17}~{\rm GeV/cm^3}$. The (anti)screening effect due to back-action can be important if
\begin{align}
    m_\varphi^2 \Lambda_\gamma^2  \lesssim \frac{2 \hbar^3}{c} \rho_\gamma\,.
    \label{eq:screening-limit}
\end{align}

Based on the above expression\,\eqref{eq:screening-limit}, for values of $m_\varphi\Lambda_\gamma$ above the long-dashed blue line in Fig.\,\ref{Fig:sensitivity-clocks} (anti)screening effects can be important and cover the entire range of parameter space accessible with a network of current optical atomic clocks. Here $\rho_\gamma$ for Earth's interior is used since this is the relevant value for the long-wavelength interactions considered. Thus sensitivity to positive-signed quadratic scalar interactions [as in Eq.~\eqref{eq:effective-scalar-mass}] is significantly diminished near Earth's surface. On the other hand, a space-based clock \cite{schiller2007optical,Schkolnik2022Apr} (or atom interferometer \cite{arvanitaki2018search}) network could largely avoid such screening effects due to the significantly lower average mass density in the interplanetary medium. In the case of negative-signed quadratic scalar interactions [as in Eq.~\eqref{eq:effective-scalar-mass}], there is antiscreening and the field amplitude can be significantly enhanced \cite{hees2018violation}. However, further analysis is required to calculate the form of the signal from stochastically fluctuating UBDM in this scenario. Finally, we note that constraints on coupling constants based on screening and antiscreening of scalar fields can be derived from experiments testing the equivalence principle \cite{hees2018violation}.



\section{Heuristic argument for the difference in sensitivity between magnetometers and clocks }
\label{app:SensiticityDifference}

As can be seen in Figs.\,\ref{Fig:sensitivity-GNOME} and \ref{Fig:sensitivity-clocks}, the sensitivity of an optical clock network to the coupling parameters $1/\Lambda_\gamma$ or $1/\Lambda_e$ far exceeds that of magnetometer networks to the equivalent coupling parameter $1/f_q$. 
Here we offer a heuristic explanation for this difference between the two cases. 
In the case of the scalar interaction [described by Eq.\,\eqref{eq:scalar-Lagrangian}] probed by clocks, the UBDM field modulates the electromagnetic binding energy of the atom [$\sim e^2/(2a_B) \approx \alpha^2 m_e c^2/2$, where $e$ is the electron charge magnitude, $a_B$ is the Bohr radius, and $m_e$ is the electron mass]. 
This effect leads to an energy shift approximately given by
\begin{align}
    \Delta E\ts{clock} \sim \prn{\frac{\hbar c \varphi_s^2}{\Lambda_{\gamma ,e}^2}} \times \prn{\frac{1}{2} \alpha^2 m_e c^2}~,
    \label{eq:clock-energy-shift-order-of-mag}
\end{align}
where the first factor in parentheses describes the effective coupling constant associated with the UBDM field (proportional to the UBDM-induced fractional energy shift) and the second factor in the parentheses describes the scale of the electromagnetic binding energy of the atom. 
This manifestation of the scalar coupling is seen from the fact that it results in an effective modulation of the fine-structure constant $\alpha$and/or electron mass $m_e$ as described by Eqs.\,\eqref{eq:scalar-mass-variation} and \eqref{eq:scalar-alpha-variation}. 
In the case of the pseudoscalar interaction Eq.\,\eqref{eq:quadratic-Hamiltonian} probed by magnetometers, it is the energy associated with the field gradient ($\sim m_\varphi c v_0$) that is modulated, leading to an energy shift approximately given by [see Eq.\eqref{eq:quadratic-Hamiltonian}] 
\begin{align}
    \Delta E\ts{mag} \sim \prn{\frac{\hbar c \varphi_s^2}{f_q^2}} \times \prn{m_\varphi c^2 \frac{v_0}{c}}\,.
    \label{eq:mag-energy-shift-order-of-mag}
\end{align}
By comparing Eqs.\,\eqref{eq:clock-energy-shift-order-of-mag} and \eqref{eq:mag-energy-shift-order-of-mag}, it is evident that the scale of the fractional energy modulation given by the first numerator in parentheses is similar in the two cases. 
Taking the ratio between the two energy shifts yields
\begin{align}
\frac{\Delta E\ts{clock}}{\Delta E\ts{mag}} \sim \frac{f_q^2}{\Lambda_\gamma^2} \frac{\alpha^2}{2}  \frac{c}{v_0} \frac{m_e}{m_\varphi}~,
\end{align}
and thus the ratio of the squares of the coupling constants determined by given energy shift measurements is
\begin{align}
\frac{\Lambda_\gamma^2}{f_q^2} \sim \frac{\Delta E\ts{mag}}{\Delta E\ts{clock}} \frac{\alpha^2}{2}  \frac{c}{v_0} \frac{m_e}{m_\varphi} \sim 0.1 \frac{\Delta E\ts{mag}}{\Delta E\ts{clock}} \frac{m_e}{m_\varphi}~.
\end{align}\\
For $m_\varphi c^2 \sim 10^{-14}~{\rm eV}$, $\Delta E\ts{mag} \sim 10^{-19}~{\rm eV}$ (corresponding to a magnetic field measurement at the level of a few pT for a nuclear spin), $\Delta E\ts{clock} \sim 10^{-16}~{\rm eV}$ (corresponding to a fractional uncertainty in a measurement of an optical transition at the $10^{-16}$ level) and $c/v_0 \sim 10^3$, we see that $\Lambda_\gamma^2/f_q^2 \sim 10^{15}$. This estimate demonstrates that the source of the enhanced sensitivity of clocks is from the large $m_e/m_\varphi$~mass ratio, expected for UBDM ($m_\varphi \ll 1$~eV/$c^2$).\\

\end{document}